\documentclass[floatfix,aps,twocolumn,preprintnumbers,amsmath,amssymb,nofootinbib,superscriptaddress,showkeys,showpacs]{revtex4}

\usepackage{graphicx,color}
\usepackage{amsmath,amssymb,bm}
\usepackage{dcolumn}

\newcommand{\np}{{\bf p}}
\newcommand{\nq}{{\bf q}}

\def\XXint#1#2#3{{\setbox0=\hbox{$#1{#2#3}{\int}$}
     \vcenter{\hbox{$#2#3$}}\kern-.5\wd0}}

\def\1{\'{\i}}

\begin{document}

\title{Global superscaling analysis of quasielastic electron scattering 
with relativistic effective mass }

\author{J.E. Amaro}\email{amaro@ugr.es} \affiliation{Departamento de
  F\'{\i}sica At\'omica, Molecular y Nuclear \\ and Instituto Carlos I
  de F{\'\i}sica Te\'orica y Computacional \\ Universidad de Granada,
  E-18071 Granada, Spain.}

\author{V.L. Martinez-Consentino}\email{victormc@ugr.es} 
  \affiliation{Departamento de
  F\'{\i}sica At\'omica, Molecular y Nuclear \\ and Instituto Carlos I
  de F{\'\i}sica Te\'orica y Computacional \\ Universidad de Granada,
  E-18071 Granada, Spain.}
  
\author{E. Ruiz
  Arriola}\email{earriola@ugr.es} \affiliation{Departamento de
  F\'{\i}sica At\'omica, Molecular y Nuclear \\ and Instituto Carlos I
  de F{\'\i}sica Te\'orica y Computacional \\ Universidad de Granada,
  E-18071 Granada, Spain.} 

\author{I. Ruiz Simo}\email{ruizsig@ugr.es} \affiliation{Departamento de
  F\'{\i}sica At\'omica, Molecular y Nuclear \\ and Instituto Carlos I
  de F{\'\i}sica Te\'orica y Computacional \\ Universidad de Granada,
  E-18071 Granada, Spain.}

\date{\today}

\begin{abstract}
\rule{0ex}{3ex} 

We present a global analysis of the inclusive quasielastic electron
scattering data with a superscaling approach with relativistic effective mass.
The SuSAM* model exploits the approximation of factorization of the scaling
function $f^*(\psi^*)$ out of the cross section under quasifree
conditions.  Our approach is based on the relativistic mean field theory of
nuclear matter where a relativistic effective mass for the nucleon
encodes the dynamics of nucleons moving in presence of scalar and
vector potentials. Both the scaling variable $\psi^*$ and the single
nucleon cross sections include the effective mass as a parameter to be
fitted to the data alongside the Fermi momentum $k_F$. Several methods
to extract the scaling function and its uncertainty from the data are
proposed and compared.  The model predictions for the quasielastic
cross section and the theoretical error bands are presented and
discussed for nuclei along the periodic table from $A=2$ to
$A=238$: 
$^2$H,  $^3$H, $^3$He, $^4$He, $^{ 12 }$C, $^{ 6 }$Li,   
$^{9  }$Be,   $^{ 24 }$Mg,   $^{59  }$Ni, 
  $^{89  }$Y,   $^{ 119 }$Sn,   
$^{181  }$Ta,   $^{186  }$W,   $^{197  }$Au,
$^{16}$O,
$^{27}$Al, $^{40}$Ca, $^{48}$Ca, $^{56}$Fe, $^{208}$Pb, and $^{238}$U.
 We find that more than 9000 of the total $\sim 20000$ data fall
within the quasielastic theoretical bands.
Predictions for $^{48}$Ti and $^{40}$Ar are also provided for the
kinematics of interest to neutrino experiments.

\end{abstract}



\maketitle

\section{Introduction}

Inclusive electron scattering is a powerful tool to study the
quasielastic response of nuclei, which arises in the region of
energy-momentum transfer, $(\omega,q)$, dominated by direct knockout
of bound nucleons. These reactions have experienced a revival due to
the recent neutrino oscillation experiments, which need precise
modeling of neutrino scattering from nuclei at intermediate energies
\cite{Mos16,Kat17,Alv14,Ank17,Ben17}. The recent measurements of CC
neutrino and antineutrino cross sections
\cite{Nomad09,Agu10,Agu13,Fio13,Abe13,Abe16,Abe18} have allowed
to test the current models as applied to neutrino scattering.
Systematic differences  between the theoretical predictions of
the neutrino and antineutrino data from different groups have been found
\cite{Mar09,Nie11,Gal16,Meg16,Meg14,Ank15,Gra13,Pan16,Mar16}.
Work is in progress to conciliate the models, 
trying to find the origin of their differences
and  to reduce the systematic errors.

However, in the neutrino experiments the incident energy cannot be
fixed, and the measurements are cross sections averages weighted by
some known neutrino flux. Therefore the detailed differences
between models of the quasielastic response of nuclei should be
further investigated through the corresponding predictions for
$(e,e')$ data.  Despite the progress achieved with nuclear models
based on first principles \cite{Lov16}, the nuclear shell model
\cite{Pan15}, or the spectral function theory \cite{Ank15b,Roc16}, the
high energies and momenta for the kinematics of interest $q \sim 1$
GeV/c require important relativistic corrections \cite{Ama02,Ama05}
that are not easy to implement in models of finite nuclei.  Other
fundamental requirements like gauge invariance or off-shell
extrapolations of the currents can also generate theoretical
ambiguities and discrepancies in the results. Moreover reaction
mechanisms modifying the impulse approximation, such as final-state
interactions, short-range correlations, meson-exchange currents, pion
emission and inelastic excitations, make not easy to construct a
sensible model providing a complete description of the whole set of
$(e,e')$ data at the full range of kinematics.

Relativity not only plays a role in the kinematics and in the current
operator, but also in the dynamics.  In a fully relativistic mean
field model \cite{Cab07} the scalar and vector relativistic potentials
enlarge the lower components of the (Dirac) nucleon wave functions in
the medium \cite{Udi99}, and this produces a notable enhancement of
the transverse response function.  This genuine relativistic dynamical
effect does not appear in semi-relativistic approximations based in
two-components (Pauli) spinors \cite{Ama05}.  Thus, the so called
enhancement of the transverse response \cite{Bod14,Gal16} cannot be
attributed fully to multi-nucleon processes with meson-exchange
currents but also, and significantly, to relativity.  This shows that
the separation of 1p-1h and 2p-2h channels in inclusive scattering is
in fact ambiguous and model-dependent.  Another example of ambiguity
due to the medium in the channel expansion appears is the $\Delta$
peak, which usually is attributed to pion emission with an
intermediate $\Delta$ resonance, but the $\Delta$ is dressed in the
medium and part of its width is produced by decay into the 2p-1h
channel \cite{Gil97,Sim16,Nie17} without pions.

Scaling studies are promising phenomenological alternatives to study
the nuclear response \cite{Alb88}.  In the superscaling approach
(SuSA) \cite{Don99a,Don99b,Mai02,Ama04} the longitudinal response
function is divided by a single-nucleon structure function and plotted
against a scaling variable $\psi'$, which is proportional to the
minimum initial momentum of the nucleons ejected by given momentum and energy
transfer $(q,\omega)$.  The scaling variable $\psi'$ is made
dimensionless by using units of the Fermi momentum $k_F$ and takes into
account the separation energy by subtraction of a parameter
$\epsilon_B$ to the energy transfer $\omega$.  The data, scaled in this
way for different kinematics and nuclear species, are found to collapse
into a universal longitudinal scaling function $f_{L}(\psi')$.  The
corresponding transverse scaling function $f_T(\psi')$ could not be
directly extracted from the data because the transverse response in
contaminated from non-quasielastic processes explicitly breaking scaling such as
pion emission and multinucleon emission originated mainly by meson
exchange currents.  In the SuSA approach it was assumed that
$f_T=f_L$ \cite{Ama04} and this allowed to construct a simple
model to predict neutrino cross sections from the $(e,e')$
data.

Even with the appropriate relativistic corrections in the kinematics
and currents \cite{Ama05} most of the nuclear models give $f_T \simeq f_L$ in
the impulse approximation.  Therefore these models ---SuSA included---
cannot describe the $(e,e')$ data without additional transverse
enhancement mechanisms.  However, including relativity in the dynamics
one naturally finds an enhancement $f_T > f_L$, going into the right
direction for the conciliation with data.  This is the case of the
relativistic mean field (RMF) model of finite nuclei \cite{Cab07},
which is based on the Dirac-Hartree theory of ref. \cite{Hor91}.  

The key ingredient to perform the upgrade SuSA-v2 \cite{Gon14} was to
include nuclear effects theoretically-inspired by the RMF.  SuSA-v2
uses an enhanced transverse scaling function $f_T$ different from
$f_L$. The new transverse scaling function was fitted to $(e,e')$
cross section data in a model including also 2p-2h MEC and inelastic
contributions \cite{Meg16b}.  An additional dependence of $f_T(\psi')$
on the momentum transfer $q$ was needed to reproduce the data.
Therefore the SuSA-v2 results violate scaling, although the model
keeps the word 'scaling' by tradition.

In recent work we have revisited the scaling approach by introducing a new
scaling function $f^*(\psi^*)$ including dynamical relativistic
effects \cite{Ama15,Ama17,Mar17} through the introduction of an
effective mass into its definition.  The resulting superscaling
approach with relativistic effective mass (SuSAM*) model describes a
large amount of the electron data lying inside a phenomenological
quasielastic band, and it has been extended recently to the neutrino
and antineutrino sector \cite{Rui18} with success and a fair agreement with
the data. SuSAM*  was first developed from the set of 
$^{12}$C data \cite{Ama15,Ama17}
and later applied to other nuclei in \cite{Mar17}.

The novel point of view of SuSAM* stems from the observation that a
large subset of $(e,e')$ data collapse into a thick band which can be
parameterized as a QE central value $f^*(\psi^*)$ plus/minus a
theoretical uncertainty. This phenomenological QE scaling band emerges
as the set of selected data which can be considered approximately
quasielastic except for interaction effects which break scaling just
by a little amount.  The success to describe the cross section data
with only one scaling function is due to the proved good
properties of the relativistic mean field, which already includes by
construction the transverse response
enhancement \cite{Ros80,Ser86,Weh93}.
Moreover, the  new phenomenological scaling function 
approximately encloses the  
universal scaling
function of the relativistic Fermi gas
\begin{equation}  
f_{\rm RFG}(\psi^*) = \frac{3}{4}(1-\psi^*{}^2) \theta(1-\psi^*{}^2)
\label{f}
\end{equation}
This is so because the mean field theory of nuclear matter already
provides a reasonable description of the quasielastic response
function \cite{Ros80,Ser86,Weh93}.  The phenomenological SuSAM*
scaling function differs from this parabolic shape and it can be
parameterized as a sum of two Gaussians.  An additional advantage of
the SuSAM* is that it keeps gauge invariance.  The original SuSA
violates this fundamental symmetry because it introduces an energy
shift to account for separation energy,  and hence initial and final
states have a different mass, presumably modifying the vertex
function.  In our case the energy shift is accounted for by the
effective mass, typically of $M^* = m^*_M/m_N \sim 0.8$ for medium-size
nuclei.

The goal of the present work is two-fold. First, to perform a global
simultaneous fit of the SuSAM* parameters to all the available data on
the $(e,e')$ database compiled in
refs. \cite{archive,archive2,Ben08}. Second, to present in
a comprehensive way  the model description of the cross section for all
the nuclei included in the fit. We also analyze in more detail several
nuclei, and compare the various prescriptions used to extract the scaling
function.  Finally we examine the predictions of our model to new
$(e,e')$ measurements for the nuclei $^{48}$Ti \cite{Dai18} 
and $^{40}$Ar, at present of
interest for current neutrino experiments.

The scheme of the paper is as follows.
 In Sect. II we present the formalism for the $(e,e')$ cross
section. In sect. III we present the results for the SuSAM*
scaling function.  In Sect. IV we present the predictions for the
cross sections of the different nuclei. In Sect. V we draw our
conclusions.

\begin{table*}
\begin{tabular}{cccccccccccc}\hline
&   & $a_1$ & $a_2$ & $a_3$ & $b_1$ & $b_2$ & $b_3$ & $c_1$ & $c_2$   \\ \hline
& central
& -0.0465
& 0.469
& 0.633
& 0.707
& 1.073
& 0.202
& -
& -
\\
Band A
& min
& -0.0270
& 0.442
& 0.598
& 0.967
& 0.705
& 0.149
& -
& -
\\
& max
& -0.0779
&  0.561
& 0.760
& 0.965
& 1.279
& 0.200
& -
& -
\\[3mm]
& central 
& -0.1335
& 0.4319
& 1.3885
& 0.5741
& 0.6539
& 0.6083
& 0.3405
& 2.2947
\\
Band B
& min
& 0.3075
& 0.6898
& 0.4115
& -0.0647
& 0.3145
& 0.3267
& -0.8362
& 0.0295
\\
& max
& -7.0719
&  2.4644
& 38.58
& -7.0724
&  2.4595
& 38.58
& -0.2613
& 0.2410
\\[3mm]
& central 
& -0.0537
& 0.5051
& 0.6055
& 0.7258
& 1.0102
& 0.2306
& -0.9765
& 0.1716
\\
Band C
&min
& -0.0435
&  0.4245
& 0.4940
& 0.5129
& 0.7360
& 0.2346
& -0.8549
&  0.0337
\\
& max
& -0.1192
& 0.4955
& 1.1504
& 0.7001
& 1.0939
& 0.3992
& -1.0058
& 1.9235
\\ \hline
\end{tabular}
\caption{Parameters of the central value of 
 the phenomenological scaling function, $f^*(\psi^*)$, 
 and those of the lower and upper boundaries (min and
  max, respectively) of the bands.  Band A correspond to the $^{12}$C
  fit of ref. \cite{Ama17}, Band B correspond to the twelve-nuclei fit
  of ref. \cite{Mar17}, and band C is the global fit performed in this
  work.}
\label{bandas}
\end{table*}

\begin{table*}
\begin{center}
\begin{tabular}{llllllllllll }
\hline
& \multicolumn{2}{c}{Visual fit}
& \multicolumn{2}{c}{No. points fit}
& \multicolumn{2}{c}{$\chi^2$ fit}
& \multicolumn{2}{c}{Global fit}
& \multicolumn{2}{c}{No. points}
&
\\
\hline
Nucleus & $k_F$ [MeV/c] & $M^*$ & $k_F$ & $M^*$ & $k_F$ & {$M^*$} & $k_F$ & $M^*$ & $N_{QE}$ & $N_{\rm tot}$ & $\chi^2/N'_{QE}$ \\
\hline\hline
$^2$H & 80 & 1.00 & 88 & 0.99 & 82 & 1.00 & 81 & 0.99 & 426 & 2135 & 0.372\\ 
$^3$H & 120 & 0.97 & 142 & 0.99 & 136 & 0.98 & 126 & 0.97 & 139 & 540 & 0.414\\ 
$^3$He & 140 & 0.95 & 147 & 0.96 & 130 & 0.98 & 130 & 0.96 & 794 & 2472 & 0.565\\
$^4$He & 160 & 0.90 & 180 & 0.89 & 180 & 0.86 & 159 & 0.87 & 803 & 2718 & 0.699\\
$^6$Li & 165 & 0.80 & --- & --- & 175 &  0.77 & 164 & 0.80 & 23 & 165 & 0.18\\ 
$^9$Be & 185 & 0.80 & --- & --- & 202 &  0.85 & 184 & 0.80 & 16 & 390 & 0.07\\ 
$^{12}$C & 225 & 0.80 & 226 & 0.82 & 217 & 0.80 & 212 & 0.83 & 660 & 2883 & 0.697\\ 
$^{16}$O & 230 & 0.80 & 259 & 0.84 & 250 & 0.79 & 228 & 0.80 & 48 & 126 & 0.999\\ 
$^{24}$Mg & 235 & 0.75 & --- & --- & 238 & 0.65 & 235 & 0.75 & 23 & 34 & 0.313\\ 
$^{27}$Al & 236 & 0.80 & 258 & 0.78 & 249 & 0.80 & 233 & 0.80 & 75 & 628 & 0.499\\ 
$^{40}$Ca & 240 & 0.73 & 250 & 0.73 & 236 & 0.71 & 229 & 0.74 & 616 & 1339 & 0.76\\ 
$^{48}$Ca & 247 & 0.73 & 242 & 0.75 & 237 & 0.71 & 230 & 0.75 & 728 & 1227 & 0.672\\ 
$^{56}$Fe & 238 & 0.70 & 240 & 0.79 & 241 & 0.70 & 240 & 0.72 & 485 & 2429 & 1.20\\ 
$^{59}$Ni & 235 & 0.67 & --- & --- & 238 & 0.65 & 234 & 0.68 & 27 & 37 & 0.09\\ 
$^{89}$Y & 235 & 0.65 & --- & --- & 224& 0.64 & 233 & 0.65 & 27 & 37 & 0.17\\ 
$^{119}$Sn & 235 & 0.65 & --- & --- & 232 & 0.64 & 236 & 0.66 & 24 & 34 & 0.204\\ 
$^{181}$Ta & 235 & 0.65 & --- & --- & 232 & 0.64 & 236 & 0.66 & 24 & 33 & 0.115\\
$^{186}$W & 230 & 0.77 & --- & --- & 226 & 0.76 & 231 & 0.80 & 45 & 184 & 0.6\\ 
$^{197}$Au & 240 & 0.75 & --- & --- & 238 & 0.78 & 235 & 0.74 & 30 & 96 & 0.237\\ 
$^{208}$Pb & 237 & 0.65 & 239 & 0.64 & 233 & 0.56 & 232 & 0.63 & 818 & 1714 & 1.223\\ 
$^{238}$U & 259 & 0.65 & 219 & 0.59 & 219 & 0.51 & 255 & 0.65 & 193 & 420 & 1.74\\
\hline
\end{tabular}
\end{center}
\caption{
Values of the parameters $M^*$ and $k_F$ (in MeV/c) obtained
from the different fits to the scaling band, the total number of data 
$N_{\rm tot}$, 
the number $N_{QE}$ of quasielastic points, and the $\chi^2$ 
divided by the number $N'_{QE}$ 
of quasielastic points ( with $-1< \psi^*<1$ 
in the case of $^2$H and $^{3}$He.}
\label{tabla}
\end{table*}

\begin{table}
\begin{center}
\begin{tabular}{|c|c|c|c|c|c|} \cline{1-6}
Nuclei & $(k_F)_c$ & $M^*_c$ & a & b & $\theta$ \\ \cline{1-6} $^2$H &
82.5 & 0.994 & 0.006 & 0.02 & 105\\ \cline{1-6} $^3$H & 136 & 0.98 &
0.016 & 0.014 & 145 \\ \cline{1-6} $^3$He & 125.5 & 0.988 & 0.010 &
0.038& 90 \\ \cline{1-6} $^4$He & 180 & 0.86 & 0.031 & 0.021 &
20\\ \cline{1-6} $^{12}$C & 218 & 0.8 & 0.027 & 0.044& 90 \\ \cline{1-6}
$^{16}$O & 252 & 0.79 & 0.034 & 0.063 & 90\\ \cline{1-6} $^{27}$Al & 249.5 &
0.795 & 0.029 & 0.05& 95 \\ \cline{1-6} $^{40}$Ca & 237.5 & 0.71 & 0.034 &
0.057 & 90 \\ \cline{1-6} $^{48}$Ca & 238 & 0.71 & 0.032 & 0.05& 90
\\ \cline{1-6} $^{56}$Fe & 242 & 0.705 & 0.061 & 0.031 & 0\\ \cline{1-6}
$^{208}$Pb & 233 & 0.56 &0.035 & 0.062 &90 \\ \cline{1-6} 
$^{238}$U & 221 & 0.52
& 0.027 & 0.064& 90\\ \cline{1-6}
\end{tabular}
\end{center}
\caption{Parameters of the $10\%$ confidence ellipses of the  $\chi^2$ fits.}
\label{ellipse}
\end{table}

\section{Formalism}

Here we summarize, for completeness and to fix our notation, 
 the general formalism of quasielastic electron
scattering and the relativistic mean field model of nuclear matter
\cite{Ama15}.  We assume that an incident electron transfers momentum
$\nq$ and energy $\omega$ to the nucleus, scattering to an angle $\theta$.  The
four-momentum transfer is denoted as 
$Q^2=\omega^2-q^2 <0$.  The quasielastic
cross section is written in the plane wave Born approximation 
with one-photon-exchange in terms of
the longitudinal and transverse response functions as
\begin{equation}
\frac{d\sigma}{d\Omega'd\epsilon'}
= \sigma_{\rm Mott}
(v_L R_L +  v_T  R_T).
\end{equation}
Here $\sigma_{\rm Mott}$ is the Mott cross section, 
the kinematic factors $v_L,v_T$ are defined by 
\begin{eqnarray}
v_L &=& 
\frac{Q^4}{q^4} \\
v_T &=&  
\tan^2\frac{\theta}{2}-\frac{Q^2}{2q^2}.
\end{eqnarray}
Finally
 $R_L(q,\omega)$ and $R_T(q,\omega)$ are 
the nuclear longitudinal and transverse response functions, respectively. 
The L and T responses are computed starting
with the RMF in nuclear matter \cite{Bar98}.  We consider one-particle one-hole
(1p-1h) excitations in the nuclear medium 
produced by one-body electromagnetic current operator,
such that the initial and final
nucleons have the {\em same} effective mass $m_N^*$.  
Thus, the initial nucleon
has energy $E=\sqrt{\np^2+m_N^*{}^2}$ in the mean field,
with $p$ below the  Fermi momentum, $p < k_F$.
The final
momentum of the nucleon is $\np'=\np+\nq$, and its corresponding energy is
$E'=\sqrt{\np'{}^2+m_N^*{}^2}$.  Pauli blocking implies $p' > k_F$.

 The nuclear response functions can be written in the
factorized form 
\begin{equation}
R_K  =   r_K f^*(\psi^*), \label{factorization}  
\end{equation}
for $K=L,T$.
Here $r_L$ and $r_T$ are the single-nucleon contributions to the 
response functions, 
averaged over the Fermi motion given below.
 $f^*(\psi^*)$ is the scaling function, given by Eq. \ref{f}.  It
depends only on the scaling variable $\psi^*$, which is defined as
follows.

First, it is convenient to  introduce the dimensionless variables 
\begin{eqnarray}
\lambda  &=& \omega/2m_N^*, \\
\kappa  & = & q/2m_N^*,\\
\tau & = & \kappa^2-\lambda^2, \\
\eta_F & = &  k_F/m_N^*,\\
\xi_F & = & \sqrt{1+\eta_F^2}-1,\\
\epsilon_F &=& \sqrt{1+\eta_F^2}, 
\end{eqnarray}
Note that in the SuSA formalism
these
variables are defined dividing by the nucleon mass $m_N$ instead
of $m_N^*$  \cite{Alb88}. 

Then, one computes the minimum energy for the initial nucleon that is
allowed to absorb the energy and momentum transfer $(q,\omega)$. From
energy and momentum conservation, in units of $m_N^*$ it is given by
\begin{equation}
\epsilon_0={\rm Max}
\left\{ 
       \kappa\sqrt{1+\frac{1}{\tau}}-\lambda, \epsilon_F-2\lambda
\right\},
\end{equation}

We can finally write the definition of the scaling variable as
\begin{equation}
\psi^* = \sqrt{\frac{\epsilon_0-1}{\epsilon_F-1}} {\rm sgn} (\lambda-\tau).
\end{equation}
The sign convention is such that 
$\psi^*$ is negative to the left of the quasielastic peak (defined by 
$\lambda = \tau$)
and positive on the right side.

The nucleonic contributions to the responses are
\begin{equation}
r_K = \frac{\xi_F}{m^*_N \eta_F^3 \kappa} (Z U^p_K+NU^n_K)
\label{single} 
\end{equation}
for $Z$ protons and $N$ neutrons.  The single-nucleon response
functions longitudinal and transverse, $U_L, U_T$ are computed from
the matrix elements of the electromagnetic current operator.

In this work we use the CC2 prescription of the electromagnetic current
operator \cite{For83}
\begin{equation}
J^\mu_{s's}=
\overline{u}_{s'}(\np')
\left[ 
F_1\gamma^\mu 
+F_2i\sigma^{\mu\nu}\frac{Q_\nu}{2m_N}
\right]u_{s}(\np)
\label{vector}
\end{equation}
where $F_i$ are the Pauli form factors of the nucleon, and the spinors
contain the effective mass instead of the bare nucleon mass. Therefore
 the above matrix element differs from the bare nucleon
result with $m_N^* = m_N$.  As a consequence 
the electric and magnetic form factors are modified in the medium 
according to \cite{Ama15, Bar98}:
\begin{eqnarray}
G_E^*  &=&  F_1-\tau \frac{m^*_N}{m_N} F_2 \\
G_M^*  &=& F_1+\frac{m_N^*}{m_N} F_2.  \label{GM}
\end{eqnarray}
 For the free Dirac and Pauli form
factors, $F_1$ and $F_2$, we use the Galster parameterization \cite{Gal71}.

Using the above definitions, 
 the single-nucleon response functions are given by
\begin{eqnarray}
U_L &=& \frac{\kappa^2}{\tau}
\left[ (G^*_E)^2 + \frac{(G_E^*)^2 + \tau (G_M^*)^2}{1+\tau}\Delta \right]
\\
U_T &=& 2\tau  (G_M^*)^2 + \frac{(G_E^*)^2 + \tau (G_M^*) ^2}{1+\tau}\Delta
\end{eqnarray}
Here we use the quantity $\Delta$ defined by
\begin{equation}
\Delta= \frac{\tau}{\kappa^2}\xi_F(1-\psi^*{}^2)
\left[ \kappa\sqrt{1+\frac{1}{\tau}}+\frac{\xi_F}{3}(1-\psi^*{}^2)\right].
\end{equation}
This is usually a small correction around the QE peak $-1<\psi^*<1$
because it is proportional to the small quantity $\xi_F$.

\begin{figure}
\includegraphics[width=8cm]{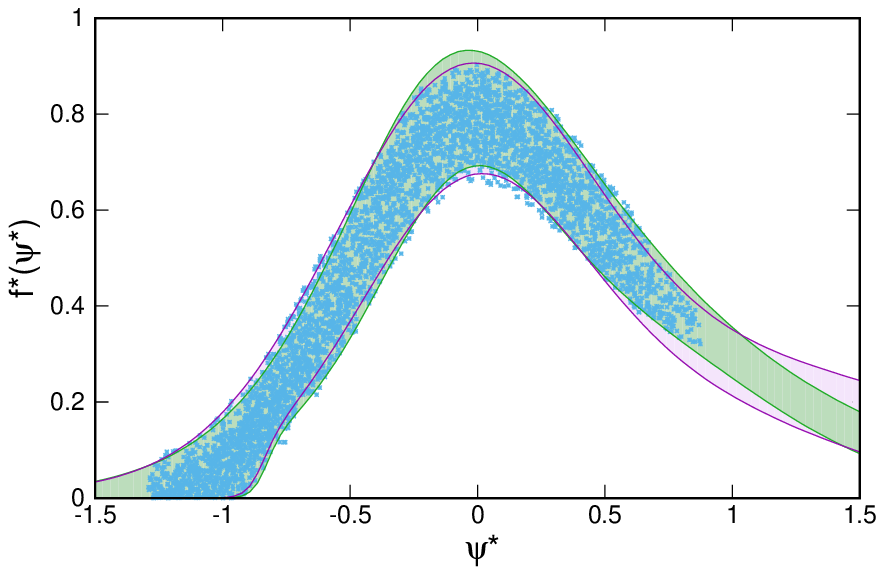}
\caption{Phenomenological scaling function bands 
compared to
  the $(e,e')$ data scaled with the best parameters
  of the global fit and selected with the density criterion.
  The  corresponding scaling function band C (in pink) is compared to
  the band B of ref. \cite{Mar17} (in green). 
  The data are selected from  \cite{Ben08,archive,archive2} }
\label{figbandas}
\end{figure}

\begin{figure*}
\includegraphics[width=\textwidth]{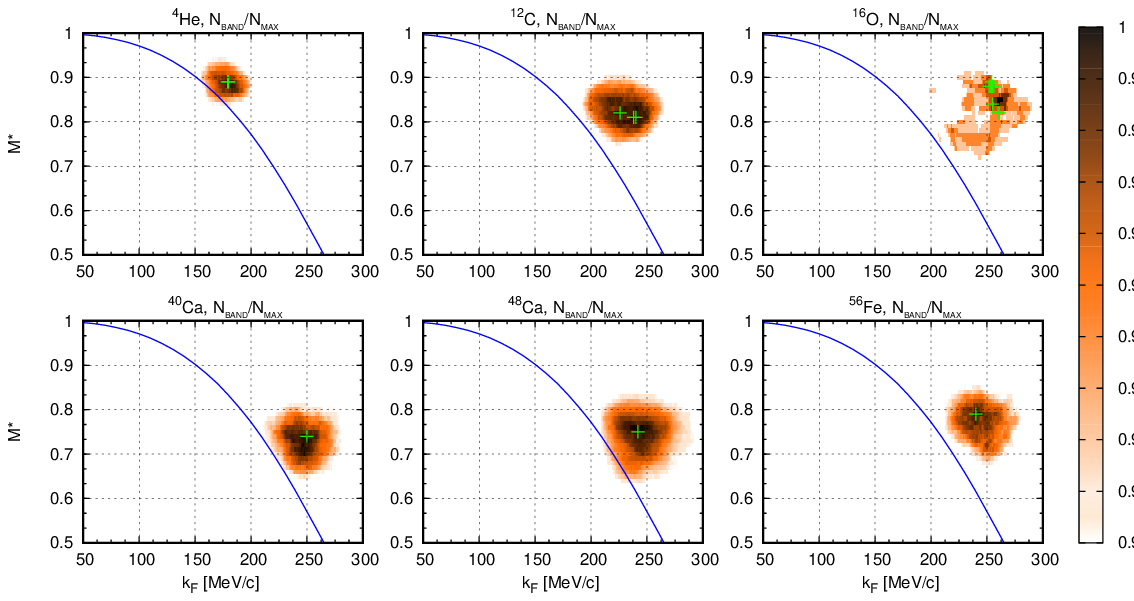}
\caption{Color maps of the number, $N$, of QE data inside the
  phenomenological band divided by $N_{\rm max}$, as a function of the
  effective mass $M^*$ and Fermi momentum $k_F$ for different
  nuclei. The $\sigma-\omega$ model of ref.  \cite{Ser86} is shown as
  comparison.  }
\label{color1}
\end{figure*}

\begin{figure*}
\includegraphics[width=\textwidth]{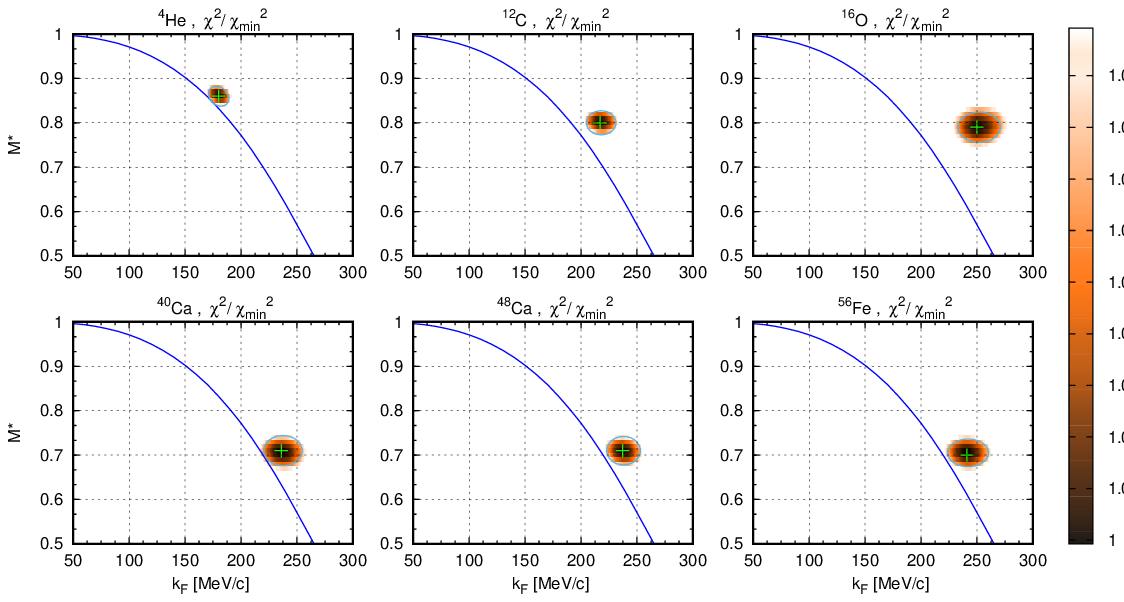}
\caption{
Color maps of the $\chi^2$  of the QE data 
 with the  phenomenological band, divided by $\chi^2_{\rm min}$, 
as a function of the
  effective mass $M^*$ and Fermi momentum $k_F$ for different
  nuclei. The region with $\chi^2$ above 10\% of the minimum is shown in white.
The region with  $\chi^2$ below 10\% of the minimum is fitted by a ellipse.
The $\sigma-\omega$ model of ref.  \cite{Ser86} is shown as
  comparison.  
 }
\label{color2}
\end{figure*}

\begin{figure*}
\includegraphics[width=\textwidth]{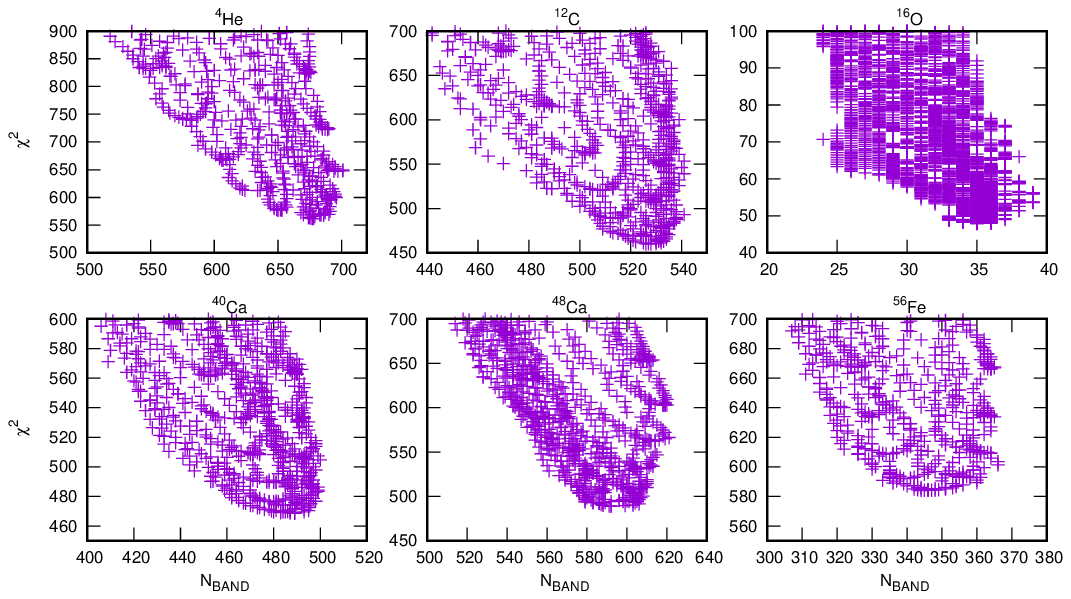}
\caption{ Correlation plot of the $\chi^2$ values versus the number of points
  inside the band $N_{\rm band}$ for different values of $k_F$ and
  $M^*$ around the region where $\chi^2$ is minimum and $N_{\rm band}$
  reaches its maximum, for six nuclei.  }
\label{maximos}
\end{figure*}

\begin{figure}
\includegraphics[width=7cm]{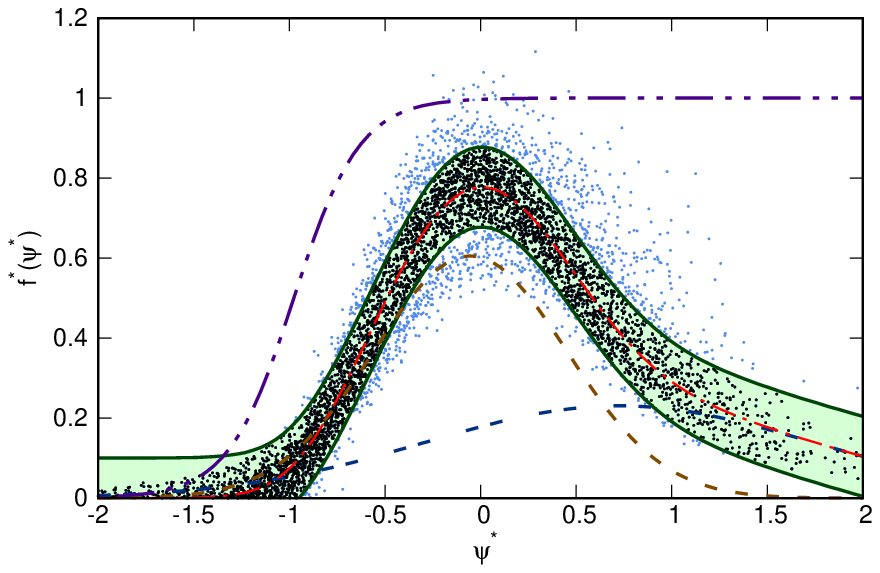}
\includegraphics[width=7cm]{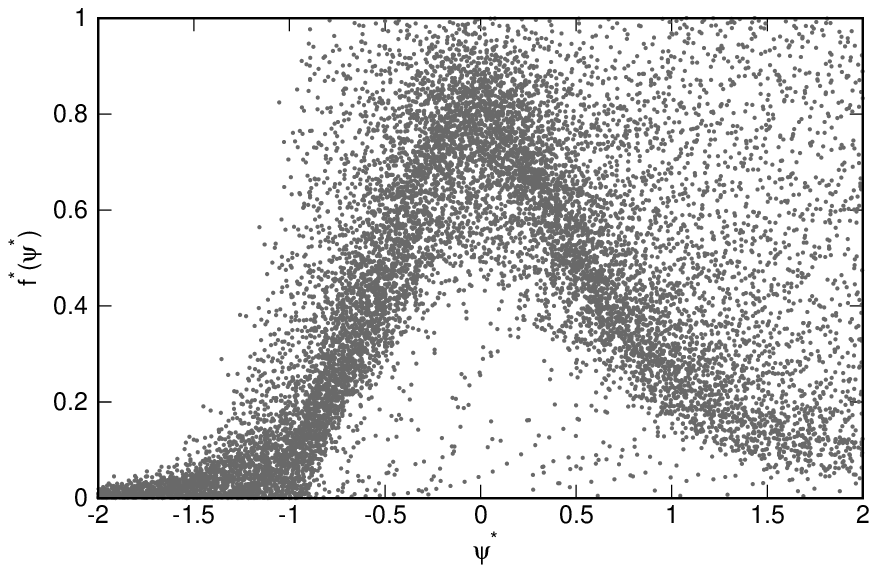}
\includegraphics[width=7cm]{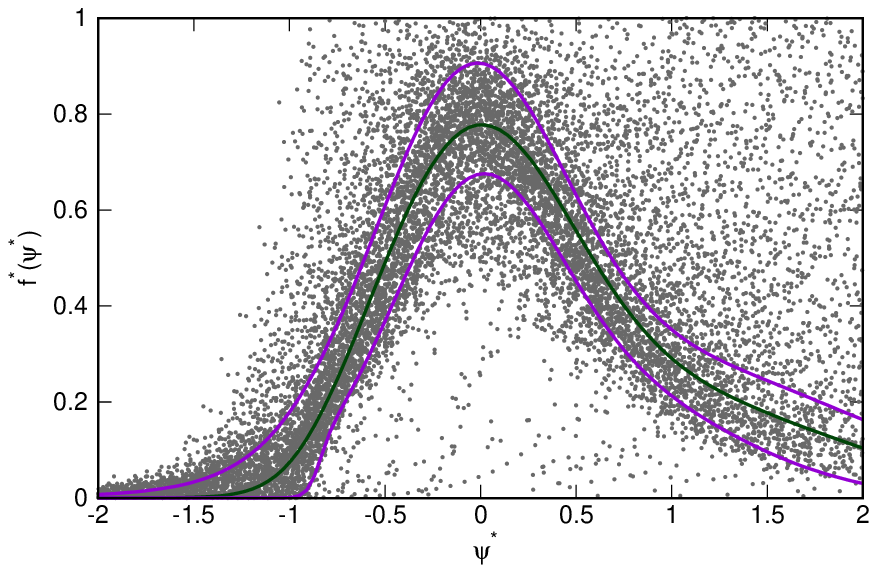}
\caption{Top panel: Data used in the global fit of the superscaling
  function.  The fit is made by maximizing the number of data inside a
  band centered around a sample scaling function of width 0.1. In
  black the 4754 points falling inside the band after the fit.  The
  central function (in red-dashed lines) 
  is the sum of two Gaussians (shown in the figure in
  dashed and double-dashed lines) modified by a Fermi function (we
  show the denominator of the Fermi function in dash-double dotted
  line). Middle panel: Global set of $(e,e')$ data scaled with the
  best parameters after the global fit.  The scaling function
  appears as a dark shadow.  Data are from ref.
  \cite{Ben08,archive,archive2}. Bottom panel: the same data 
  compared to the parameterized band C.  }
\label{datos1}
\end{figure}


\section{The SuSAM* approach}

In the SuSAM* approach the cross section is computed by using Eqs. (2,5),
 but replacing the
RFG scaling function by a phenomenological one $f^*(\psi^*)$ extracted
from the experimental data.

This can be done in several ways with a careful analysis of the
scaling properties of $(e,e')$ cross section data.  We carried out
such analyses in refs. \cite{Ama15,Ama17,Mar17}. 

In this work we extend those studies starting with $\sim 20000$
experimental $(e,e')$ cross section data for 21 nuclei:
$^2$H,  $^3$H,  $^3$He, $^4$He, $^{ 12 }$C, $^{ 6 }$Li,   
$^{9  }$Be,   $^{ 24 }$Mg,   $^{59  }$Ni, 
  $^{89  }$Y,   $^{ 119 }$Sn,   
$^{181  }$Ta,   $^{186  }$W,   $^{197  }$Au, $^{16}$O,
$^{27}$Al, $^{40}$Ca, $^{48}$Ca, $^{56}$Fe, $^{208}$Pb, and $^{238}$U.
  For
every datum we compute the corresponding experimental scaling function
$f^*_{\rm exp}$ by dividing the experimental cross section by the
single nucleon function introduced in the last section.
\begin{equation}
f^*_{\rm exp} =
\frac{\left(\frac{d\sigma}{d\Omega'd\epsilon'}\right)_{\rm exp}}{
  \sigma_{\rm Mott}\left( v_L r_L + v_T r_T \right)}
\end{equation}
 From the experimental kinematics we also compute the corresponding
 value of the scaling variable $\psi^*$.  When we plot $f^*$ versus
 $\psi^*$ one observes that a subset of data are concentrated around a
 band, but the scaling is not perfect. One then tries to change the
 values of the parameters $M^*$ and $k_F$ up to find the best scaling
 possible.

The analysis has been developed in several stages that we describe
next. For completeness, we summarize with some details what we did in
the past works \cite{Ama15,Ama17,Mar17}, 
while we will explain in depth those aspects of the
fits that were not accounted for previously. The goals of this section
are:
\begin{enumerate}
\item  To show the self consistency of the extraction of the scaling
function,
\item  to check that the different methods produce similar results for
the SuSAM* parameters, and 
\item to perform a global fit of the parameters and scaling function
  simultaneously.
\end{enumerate}

\subsection{The scaling function}

 We started with the
$^{12}$C data, by tuning the parameters $M^*$ and $k_F$ until one
finds the best scaling possible.  This was reached in \cite{Ama15} 
for $M^*=0.8$ and
$k_F=225$ MeV/c.  By applying a density criterion, a data cloud around
the RFG scaling function was selected. 
According with the criterion we  selected those scaled data 
surrounded by more than 25 data 
 inside a circle of radius $r=0.1$ in the $f^*(\psi^*)$ graph.
These selected data defined a
``QE'' region as a thick band which we parameterized as a combination
of two Gaussian functions \cite{Ama17}  
\begin{equation}
f^*(\psi^*) = 
 a_3e^{-(\psi^*-a_1)^2/(2a_2^2)}+ b_3e^{-(\psi^*-b_1)^2/(2b_2^2)}
\end{equation}
The parameters of this band A are given in Tab. \ref{bandas}.

Starting with band A we applied in ref. \cite{Mar17} several methods
to obtain the relativistic effective mass and the Fermi momentum of
all the nuclei from the periodic table, for which $(e,e')$ data
existed in our data base, taken from \cite{archive,archive2}.  With
these values of $(k_F,M^*)$ parameters we verified that all these
nuclei approximately scale similarly to the $^{12}$C ones.  These
parameters are shown in columns 2--7 of table \ref{tabla}.  For this work we
have revised the analysis of \cite{Mar17} for the nuclei $^2$H and
$^3$He, which have been updated in Tab. \ref{tabla}.

The procedure required to obtain first $k_F$ and $M^*$ for all the
nuclei from a visual fit (columns 2--3 of Tab. \ref{tabla}), providing a good
qualitative scaling of the experimental data.  With these parameters
we scaled the data for the twelve main nuclei of the data base.  
Then, we proceeded to a
more precise determination of the phenomenological scaling function 
by  discarding those kinematics where the energy transfer at
the QE peak is lower than $\sim 80-100$ MeV, and also those of high
energy were the QE peak is indistinguishable due to inelastic
dominance.

With this set of data a density criterion was newly applied,
obtaining a new SuSAM* phenomenological band.  In \cite{Mar17} that
scaling band was parameterized as the sum of two Gaussians, modified to
improve the low energy region by applying a Fermi function.
\begin{equation}
f^*(\psi^*) = \frac{
 a_3e^{-(\psi^*-a_1)^2/(2a_2^2)}+ b_3e^{-(\psi^*-b_1)^2/(2b_2^2)}}
{1+e^{-\frac{\psi^*-c_1}{c_2}}}
\label{scaling-function}
\end{equation}
The parameters of this scaling function are given in Table
\ref{bandas}.  They are labeled as 'band B'. We provide the parameters of
the lower (min), and upper (max) limits of the boundary, defining
the uncertainty band. The central parameters correspond to the best
fit to the selected data.  Band B is shown as the green band in
Fig. 1.  This is compared to the band C that is obtained in the global
fit explained below. Both bands are similar around the quasielastic
region, and therefore they are interchangeable in cross section
calculations.  Band B is the one used in the SuSAM* model of next
section to compute the QE cross sections of nuclei.

Note that the scaling band shown in Fig. 1 is well defined only in 
the quasielastic region $-1 < \psi^* <1$, while we  cannot describe 
in detail the left tail of the cross section, corresponding to $\psi^* < -1$,
and related to higher momentum components produced 
mainly by nucleon-nucleon short
range correlations \cite{Wir14, Rui16}, 
which break $M^*$-scaling. A detailed study of this 
region is beyond of the scope of the present work.

\subsection{The parameters $k_F$ and $M^*$}

We recall that band B was obtained in the last subsection from the
scaling of data using a visual fit of the parameters $k_F$ and $M^*$.
For consistency it is  pertinent to recompute those parameters
with a more quantitative procedure, which we describe next. 
This will allow us also to obtain
in return an estimate of their uncertainties. We can proceed in two
different ways, which ultimately produce 
similar results for the parameters.  The
first is to maximize the number of points inside band B. The resulting
parameters are given in columns 4 and 5 of table \ref{tabla}. The second method
is to minimize a $\chi^2$ function computed from the distances of the
data to the center of the band for each nucleus, divided by the total error, taking into account the band width
\begin{equation}
\chi^2= \sum _{i=1}^{N'_{QE}} 
\frac{(f^*(\psi^*_i)_{\rm exp}-f^*(\psi^*_i)_{\rm central})^2}
{(\Delta f^*(\psi^*_i)_{\rm exp})^2+(\Delta f^*(\psi^*_i)_{\rm th})^2}.
\end{equation}
were we have added in quadrature 
in the denominator the experimental and theoretical errors. 
The parameters resulting from minimization of this $\chi^2$ are given
in columns 6 and 7 of table \ref{tabla}.

Note that the number of points $N'_{QE}=N_{QE}$ included in the sum
has been selected by leaving only the points that are clearly around
the QE peak. These numbers of points are presented in the tenth column
of Table \ref{tabla}, together with the total number of points before
the selection, which are shown in the eleventh column of the same table.
The same data set used in the $\chi^2$ minimization has been used in
the maximization of the number of points inside the band.  However in
the case of the nuclei $^2$H and $^3$He, revised in the present work,
we have to include in the $\chi^2$ only those points with $-1 < \psi^*
< 1$ to obtain reasonable results. Therefore, for these two nuclei
$N'_{QE} < N_{QE}$.  For nine of the nuclei the number of experimental
data is not large enough to obtain a reasonable fit, and those are
hence not appearing in columns 4, 5 of table \ref{tabla}.

The values of the parameters $k_F$ and $M^*$  obtained by these
quantitative fits are similar between them, and are also similar to
the ones used in the visual fit from which band B was obtained.  The
agreement between the parameters obtained with different fits faithfully
points to the steadiness and robustness 
of the present scaling approach for this purpose.

Besides, these methods allow to compute estimations of the theoretical
errors in the parameters. The procedure is illustrated in the color
maps of Figs. 2 and 3 for six selected nuclei. 

In Fig. 2 we show the no. of data inside band B divided by the
maximum, $N_{\rm band}/N_{\rm max}$ as a function of $M^*$ and
$k_F$. We show the cloud where $N_{\rm band}/N_{\rm max}$ changes
between 0.9 and 1.  This means that changing the parameters around the
position of the maximum inside the cloud the no. of data that get out
of band B is less than 10\% of the maximum $N_{\rm max}$.  Note that
for some nuclei there is more than one maximum. In that case we
display in Tab. \ref{tabla} the values closer to the $\sigma-\omega$ theory of
the Walecka model \cite{Ser86} , also shown in the figure.  Note that
maximizing the number of points inside the band is a discretized
procedure and as consequence the shapes of the clouds deviate from a
regular elliptic shape and this does not allow to parameterize the
error in a systematic way.

The $\chi^2$ minimization method shown in Fig. 3 is more appropriate
to this end. In the figure we show the cloud in parameter space where
the $\chi^2$ values divided by the minimum $\chi^2_{\rm min}$ are in
the range 1.0--1.1. The resulting $10\%$-change clouds are more
elliptic shaped and smaller than the clouds of fig. 2. With these
plots we are able to parameterize the cloud shapes using ellipses with
three constants, $a, b, \theta$:
\begin{eqnarray}
k_F(s) & = & (k_F)_c \nonumber\\
&&   \kern -1cm   + 300 {\rm MeV/c} [ a\cos\theta \sin(s)+b\sin\theta\cos(s) ]\\
M^*(s) & = & M^*_c - a\sin\theta\sin(s) + b\cos\theta\cos(s)
\end{eqnarray}
Where $s$ is the parameter of the ellipse.
The ellipse parameters encode the errors in $k_F, M^*$ ( 10\%
confidence interval) and they are given in table \ref{ellipse}.  Note that the
centers of the ellipses are not exactly at the minimum $\chi^2$
position because we are just interested in a rough estimation 
of the error and therefore we compute 
the ellipses with a finite variation of
$10\%$ in the $\chi^2$ value.

Notice that the maximum of the number of points inside the band does
not coincide with the minimum of $\chi^2$. This is so because a set of
points inside the band can occupy different positions resulting in
different values of the $\chi^2$. Thus, the value of $\chi^2$ is not
directly related to $N_{\rm band}$, although some correlations can be
found between these two functions. The correlations between $N_{\rm
  band}$ and $\chi^2$ depend on the particular nucleus, and the
selected set of quasielastic data entering in the fit. This
correlation is shown in Fig. \ref{maximos} for six of the nuclei. In
the figure we display the values of $\chi^2$ versus $N_{\rm band}$ for
different values of $k_F$, $M^*$ around the extreme regions shown in
Figs. 2, 3. The correlation between $\chi^2$ and $N_{\rm band}$ is
stronger when $N_{\rm band}$ increases and $\chi^2$ decreases. But we
clearly observe that the maxima of $N_{\rm band}$ do not minimize
$\chi^2$, but they are close to do it.

\subsection{The global fit}

One of the goals of the present paper is to validate the universality
of the scaling function by investigating the self consistency of
the extraction method by an alternative way. In order to guarantee that
the procedure is independent on a particular nuclear species, we have
developed a global approach where the scaling function is not given
{\em a priori}. Instead we fit at the same time all the parameters of
the model, including the scaling function $f^*(\psi^*)$, and the Fermi
momentum and effective mass of all the nuclei simultaneously.  This
global fit maximizes the number of QE data points, $f^*_{\rm
  exp}(\psi^*)$, falling inside a band around a scaling function,
which we parameterize as a modified combination of Gaussians with eight
parameters, as given in Eq. (\ref{scaling-function}).  We apply the
downhill simplex method with fifty parameters (the Fermi momenta and
effective mass of 21 nuclei, $M^*, k_F$, plus the eight parameters of
the scaling function).  The ``scaling'' band used in the fit has a
constant width. It is defined by the limits $f^*(\psi^*)\pm 0.1$,
i.e., for each datum and for each set of parameters we accept the datum
inside the band if $|f^*(\psi^*)_{\rm exp} -f^*(\psi^*)|< 0.1$.  We
start with 'good' initial parameters obtained in one of the previous
separate fits.  The result of this fit is shown in the top panel of
fig. \ref{datos1}. The values of $k_F$ and $M^*$ are given in columns
8 and 9 of table \ref{tabla}. The parameters of the scaling function are given
in table \ref{bandas} as the central part of band C.

This global fit only allows to obtain the central part of the scaling
function but not the width of the band, which for the fit purposes has
been fixed to a reasonable value chosen as $\pm 0.1$, because in the
previous analyses we have seen that this is the observed order of magnitude. 
To finish the extraction of the phenomenological band we
therefore apply again a density criterion to select the true QE data
to all the QE data scaled with the global parameters. We then obtain
the set of true ``scaling'' QE data shown in fig. 1. There are 4230
points in that figure, which are a 70\% of all the points entering in
the fit. These points clearly define a band which is again
parameterized in the same way as before as in
Eq. (\ref{scaling-function}) and the parameters are given in table 1
as band C. This band is shown in pink in Fig. 1. We can see that the
result of this global fit is very similar to band B obtained by
partial fits. These results enforce the self consistency hypothesis and
the universality of the scaling function obtained.

To gain a perspective of the quality of the results we show in the
middle panel of Fig. \ref{datos1} all the data points (not only the QE
ones) for all the nuclei scaled with the parameters of the global fit.
Clearly a large fraction of data collapse into a dark shadow which
resembles to our previously determined bands. In the bottom panel of figure
\ref{datos1} the same points are compared to the global scaling
function and band C.  In fact the number of points that collapse
inside band C in fig. \ref{datos1} 
is more than 9000 of the total $\sim 20000$ data.

\section{Cross section results}

\begin{figure*}
\includegraphics[width=\textwidth]{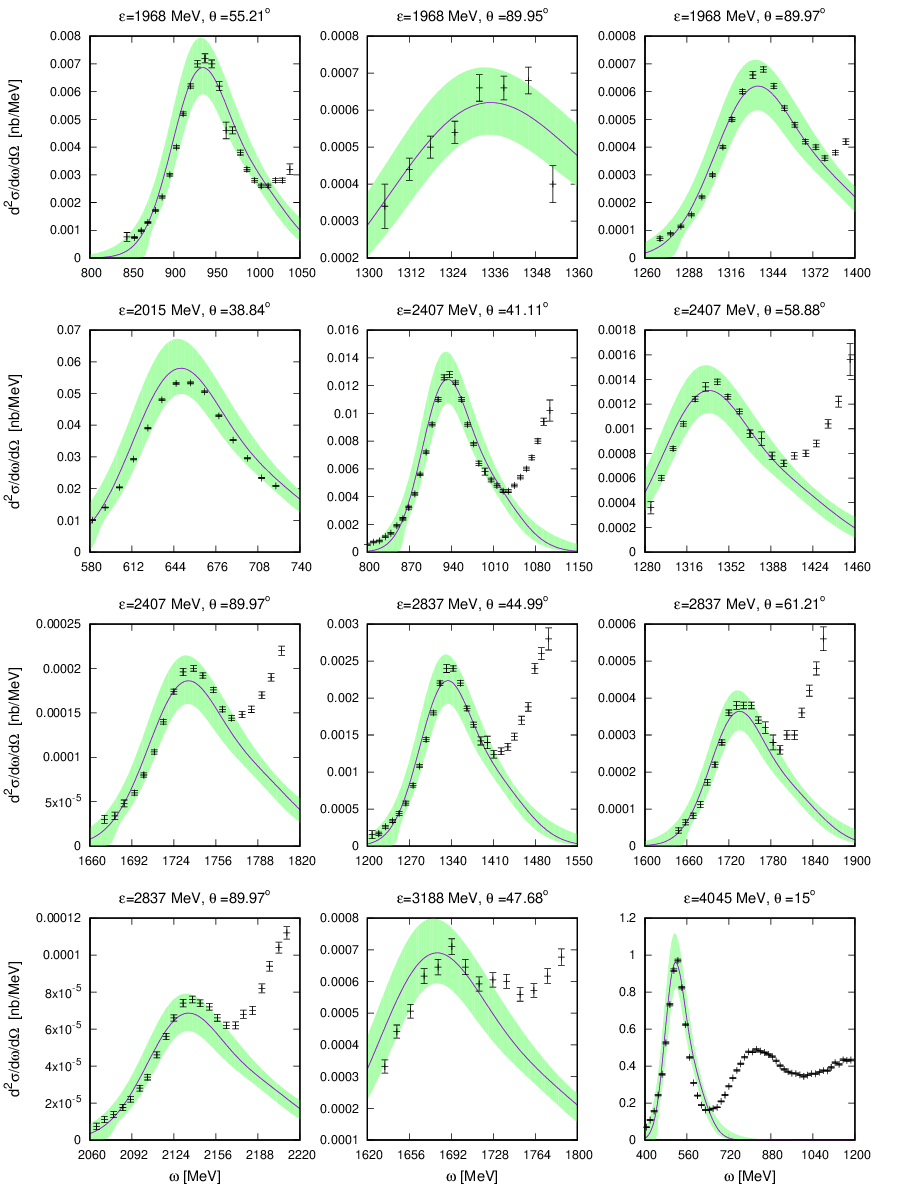}
\caption{Inclusive $(e,e')$ cross section data for $^{2}$H 
for selected kinematics compared to the SuSAM* model as a function of 
energy transfer.
 Data are from ref. 
\cite{Ben08,archive,archive2} 
}
\label{H2}
\end{figure*}

\begin{figure*}
\includegraphics[width=\textwidth]{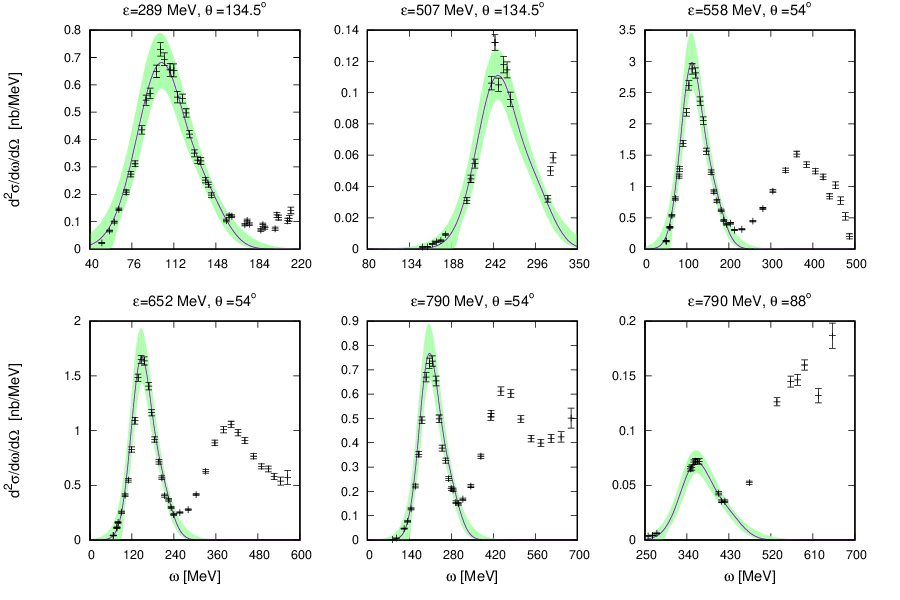}
\caption{Inclusive $(e,e')$ cross section data for $^{3}$H 
for selected kinematics compared to the SuSAM* model as a function of 
energy transfer.
 Data are from ref. 
\cite{Ben08,archive,archive2} 
}
\label{H3}
\end{figure*}

\begin{figure*}
\includegraphics[width=\textwidth]{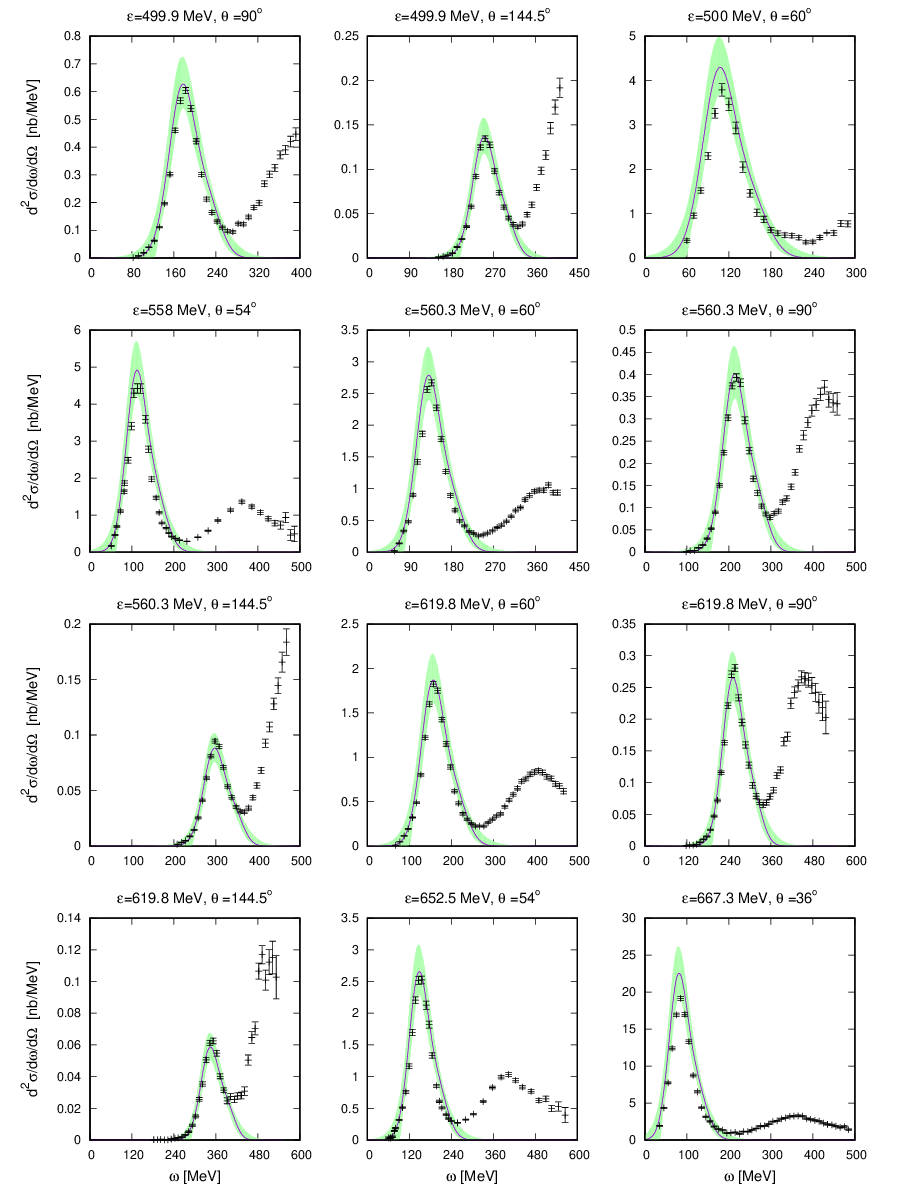}
\caption{Inclusive $(e,e')$ cross section data for $^{3}$He 
for selected kinematics compared to the SuSAM* model as a function of 
energy transfer.
 Data are from ref. 
\cite{Ben08,archive,archive2} 
}
\label{He3}
\end{figure*}

\begin{figure*}
\includegraphics[width=\textwidth]{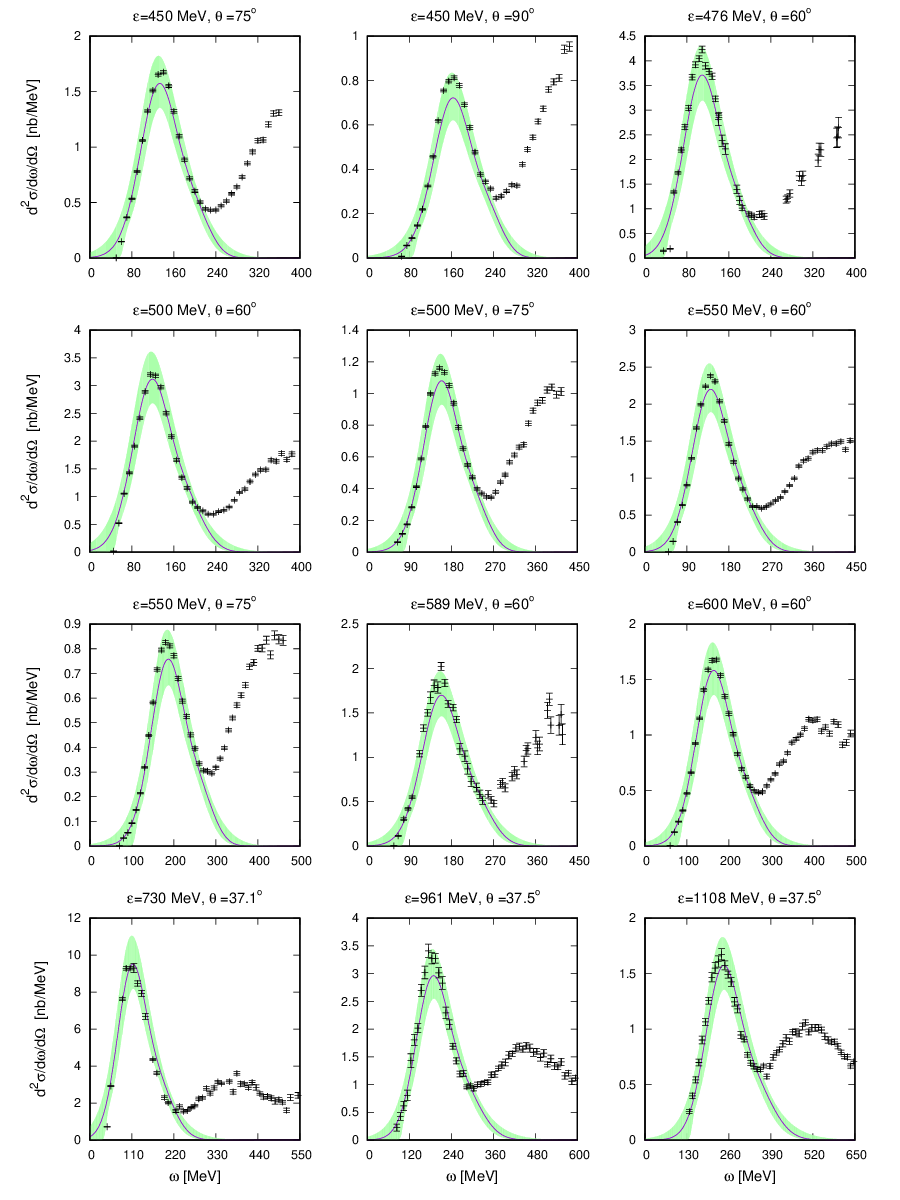}
\caption{Inclusive $(e,e')$ cross section data for $^{4}$He 
for selected kinematics compared to the SuSAM* model as a function of 
energy transfer.
 Data are from ref. 
\cite{Ben08,archive,archive2} 
}
\label{He4}
\end{figure*}

\begin{figure*}
\includegraphics[width=\textwidth]{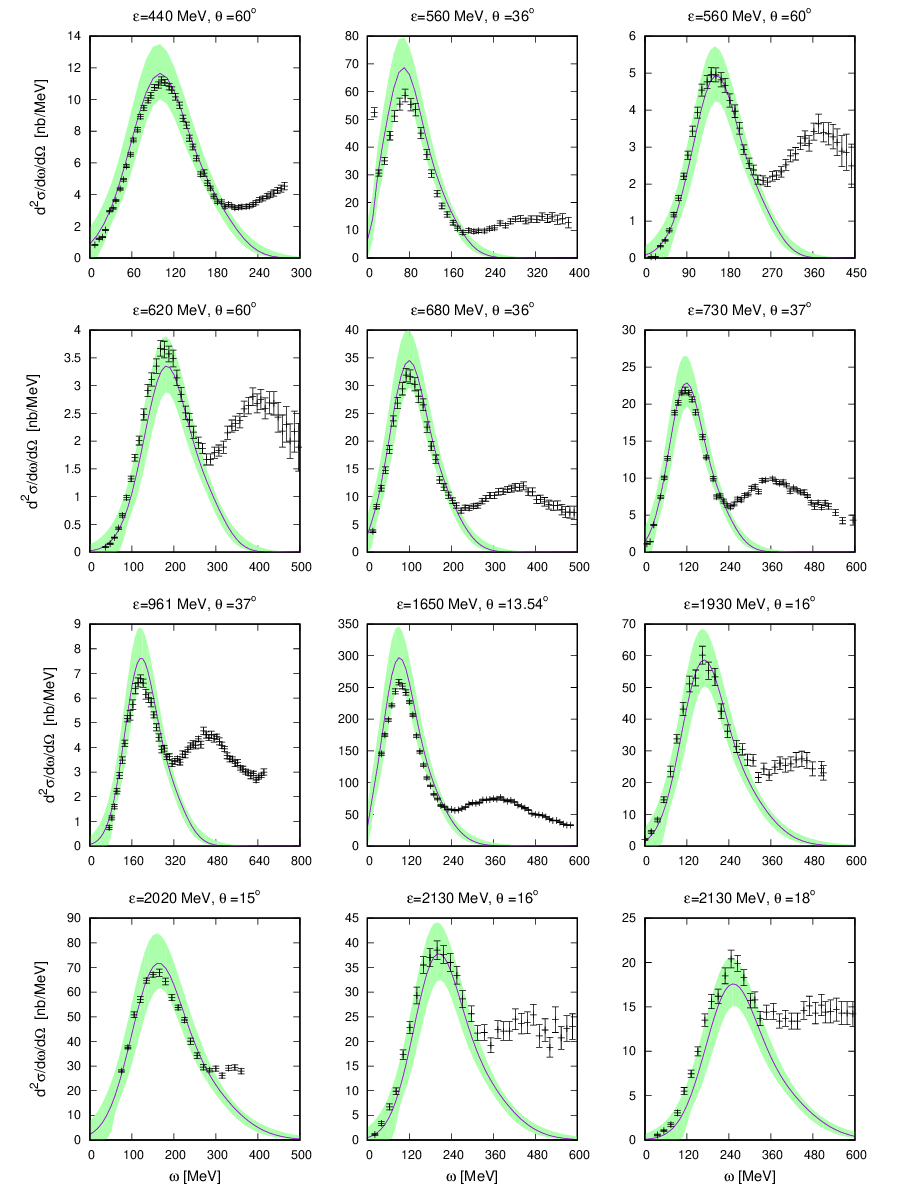}
\caption{Inclusive $(e,e')$ cross section data for $^{12}$C 
for selected kinematics compared to the SuSAM* model as a function of 
energy transfer.
 Data are from ref. 
\cite{Ben08,archive,archive2} 
}
\label{C12}
\end{figure*}

\begin{figure*}
\includegraphics[width=\textwidth]{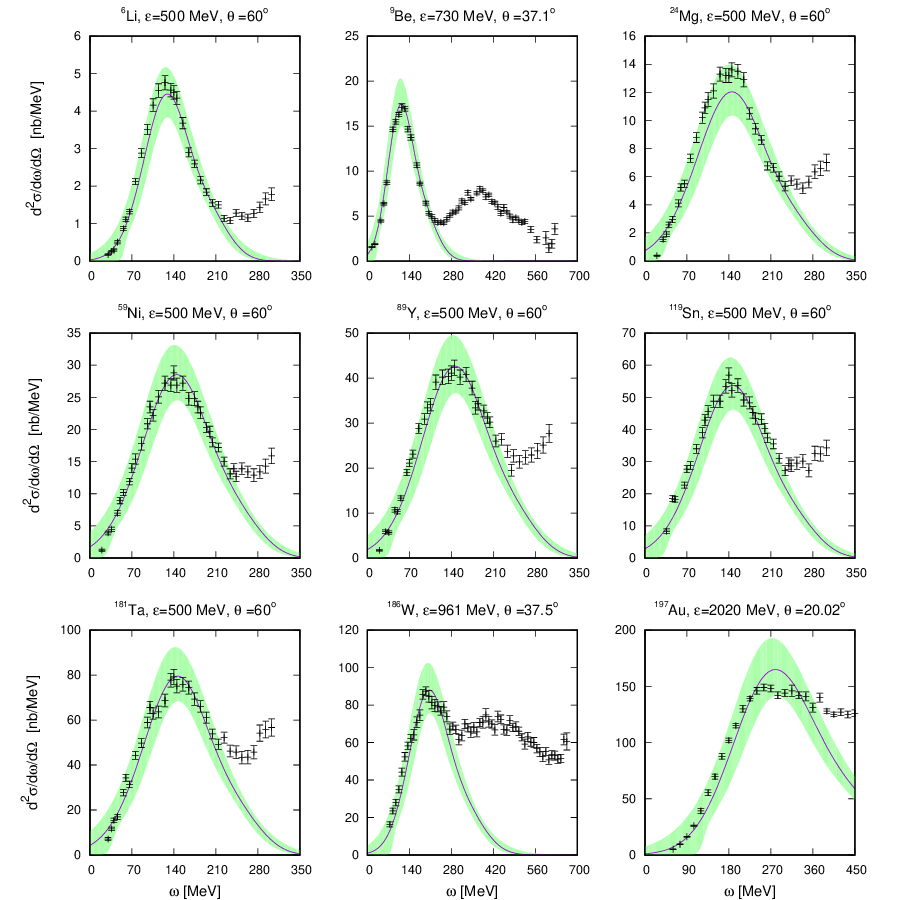}
\caption{Inclusive $(e,e')$ cross section data for nine different nuclei 
for different kinematics compared to the SuSAM* model as a function of 
energy transfer.
 Data are from ref. 
\cite{Ben08,archive,archive2} 
}
\label{nuclei}
\end{figure*}

\begin{figure*}
\includegraphics[width=\textwidth]{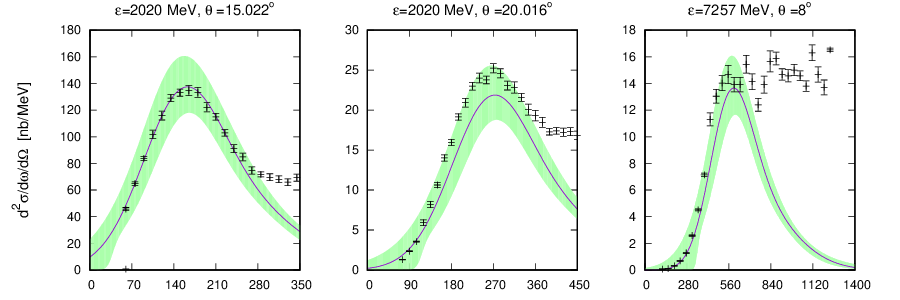}
\caption{Inclusive $(e,e')$ cross section data for $^{27}$Al 
for selected kinematics compared to the SuSAM* model as a function of 
energy transfer.
 Data are from ref. 
\cite{Ben08,archive,archive2} 
}
\label{Al27}
\end{figure*}

\begin{figure*}
\includegraphics[width=\textwidth]{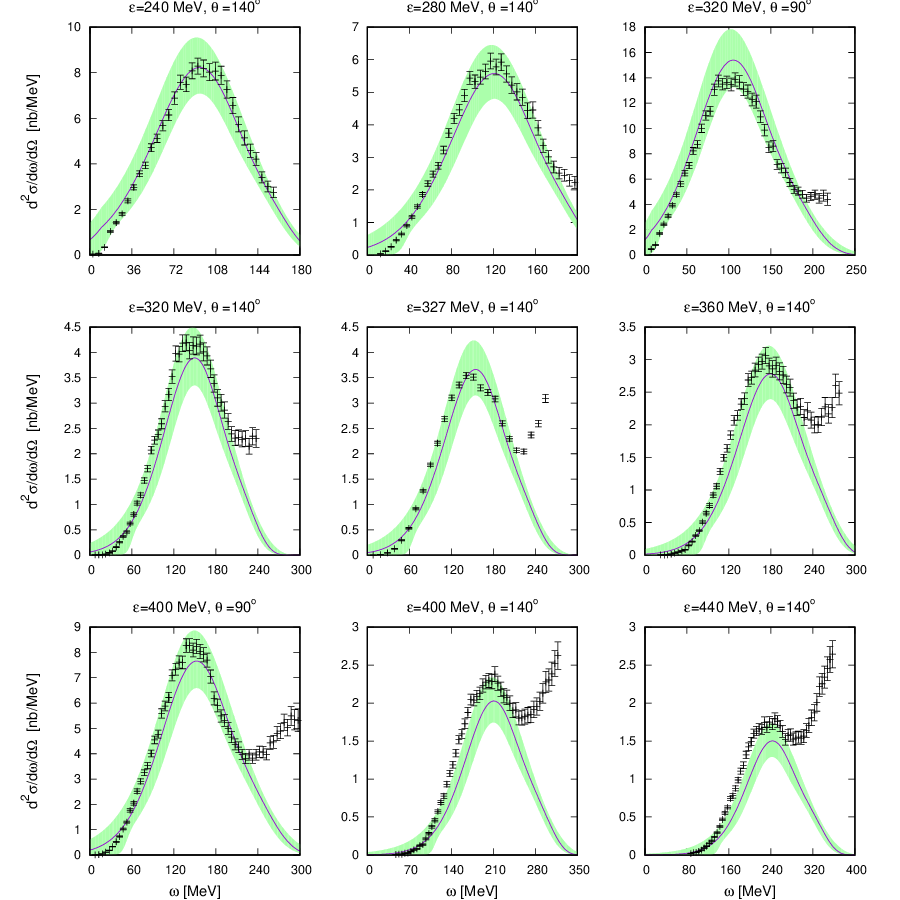}
\caption{Inclusive $(e,e')$ cross section data for $^{40}$Ca 
for selected kinematics compared to the SuSAM* model as a function of 
energy transfer.
 Data are from ref. 
\cite{Ben08,archive,archive2} 
}
\label{Ca40}
\end{figure*}

\begin{figure*}
\includegraphics[width=\textwidth]{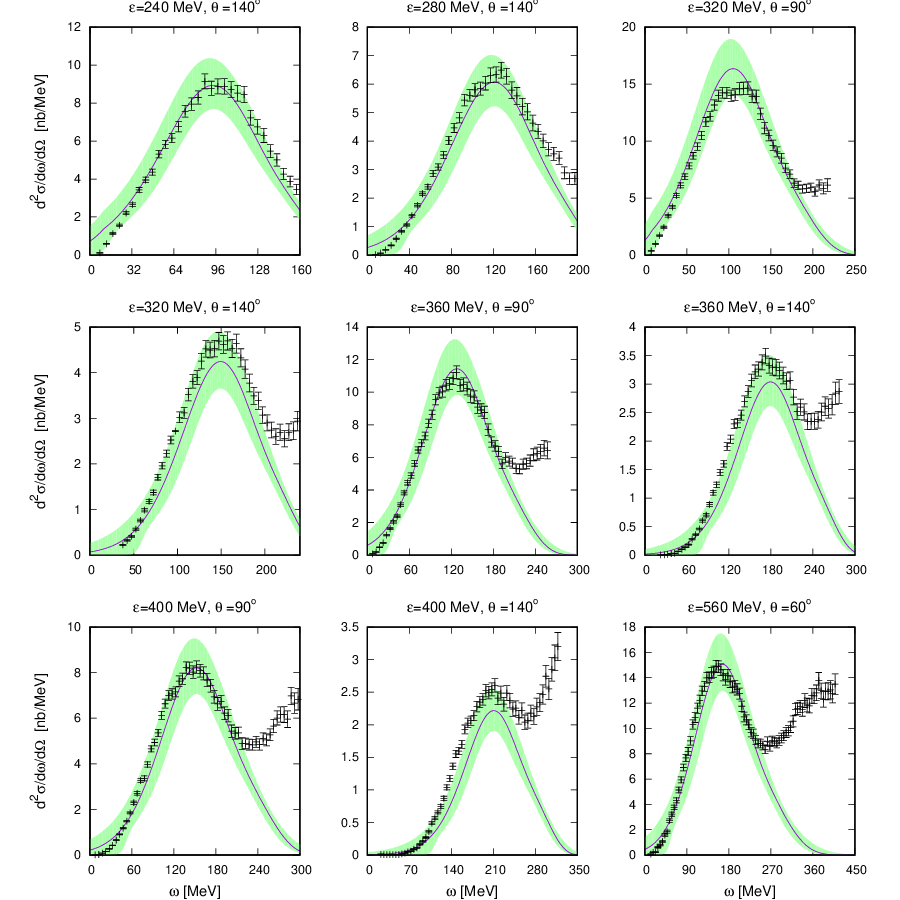}
\caption{Inclusive $(e,e')$ cross section data for $^{48}$Ca 
for selected kinematics compared to the SuSAM* model as a function of 
energy transfer.
 Data are from ref. 
\cite{Ben08,archive,archive2} 
}
\label{Ca48}
\end{figure*}

\begin{figure*}
\includegraphics[width=\textwidth]{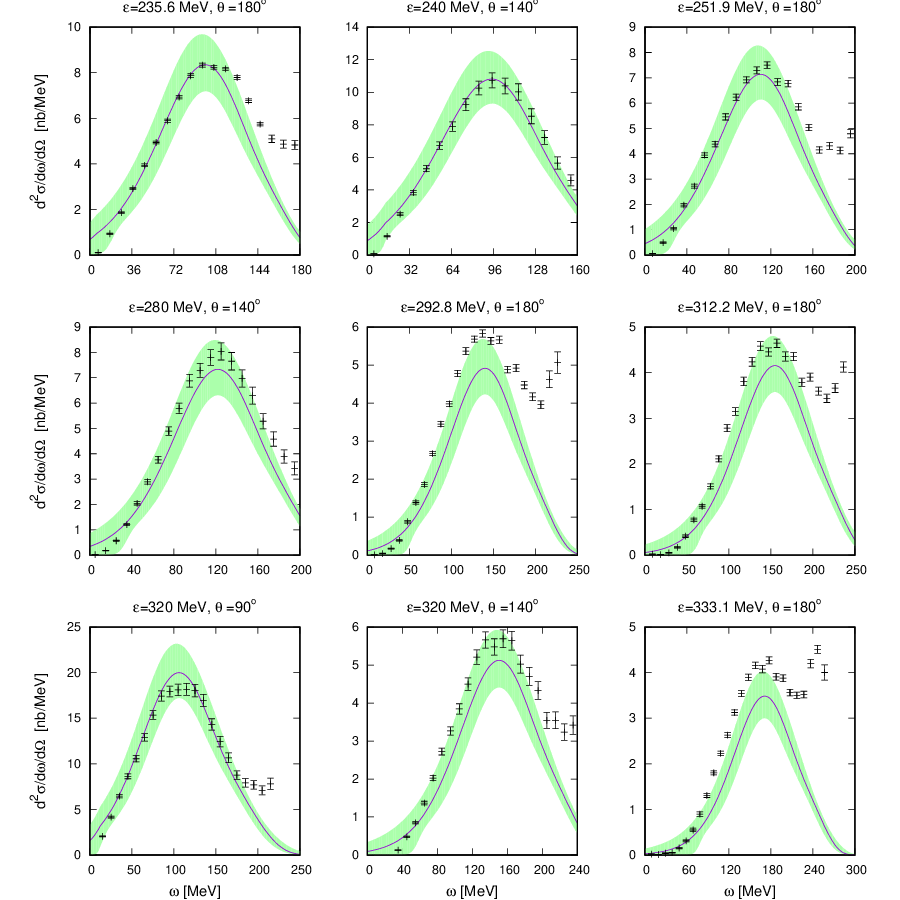}
\caption{Inclusive $(e,e')$ cross section data for $^{56}$Fe 
for selected kinematics compared to the SuSAM* model as a function of 
energy transfer.
 Data are from ref. 
\cite{Ben08,archive,archive2} 
}
\label{Fe56}
\end{figure*}

\begin{figure*}
\includegraphics[width=\textwidth]{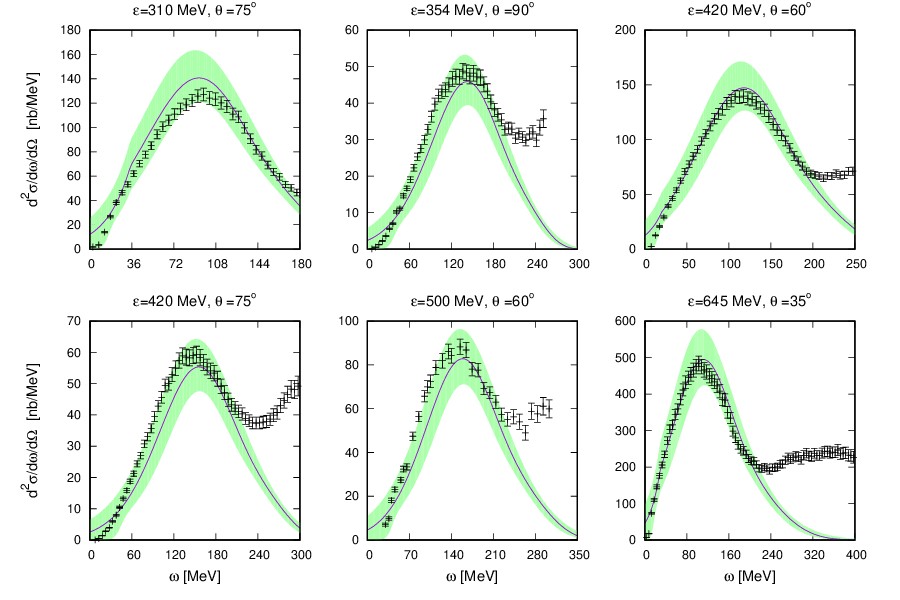}
\caption{Inclusive $(e,e')$ cross section data for $^{208}$Pb 
for selected kinematics compared to the SuSAM* model as a function of 
energy transfer.
 Data are from ref. 
\cite{Ben08,archive,archive2} 
}
\label{Pb208}
\end{figure*}

\begin{figure*}
\includegraphics[width=\textwidth]{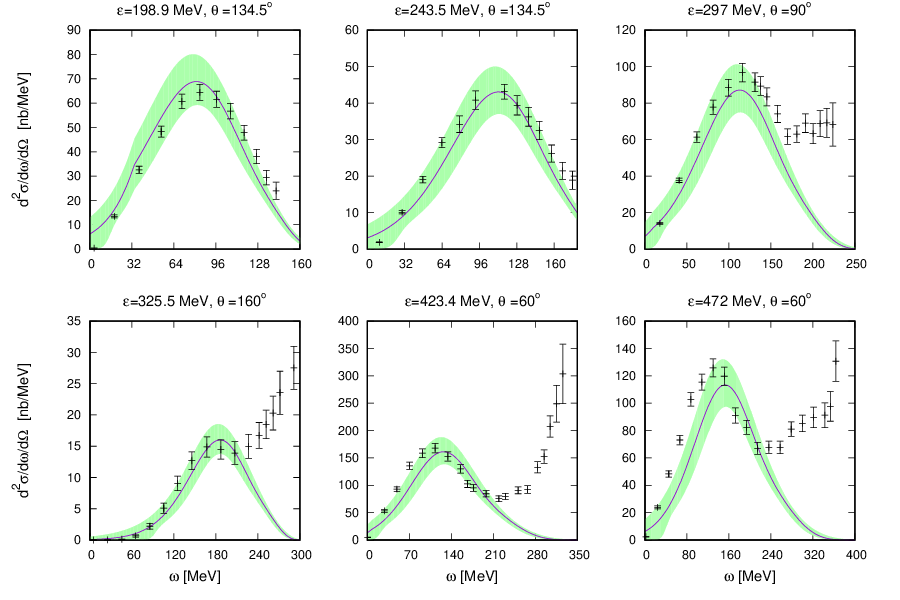}
\caption{Inclusive $(e,e')$ cross section data for $^{238}$U 
for selected kinematics compared to the SuSAM* model as a function of 
energy transfer.
 Data are from ref. 
\cite{Ben08,archive,archive2} 
}
\label{U238}
\end{figure*}

\begin{figure*}
\includegraphics[width=15cm]{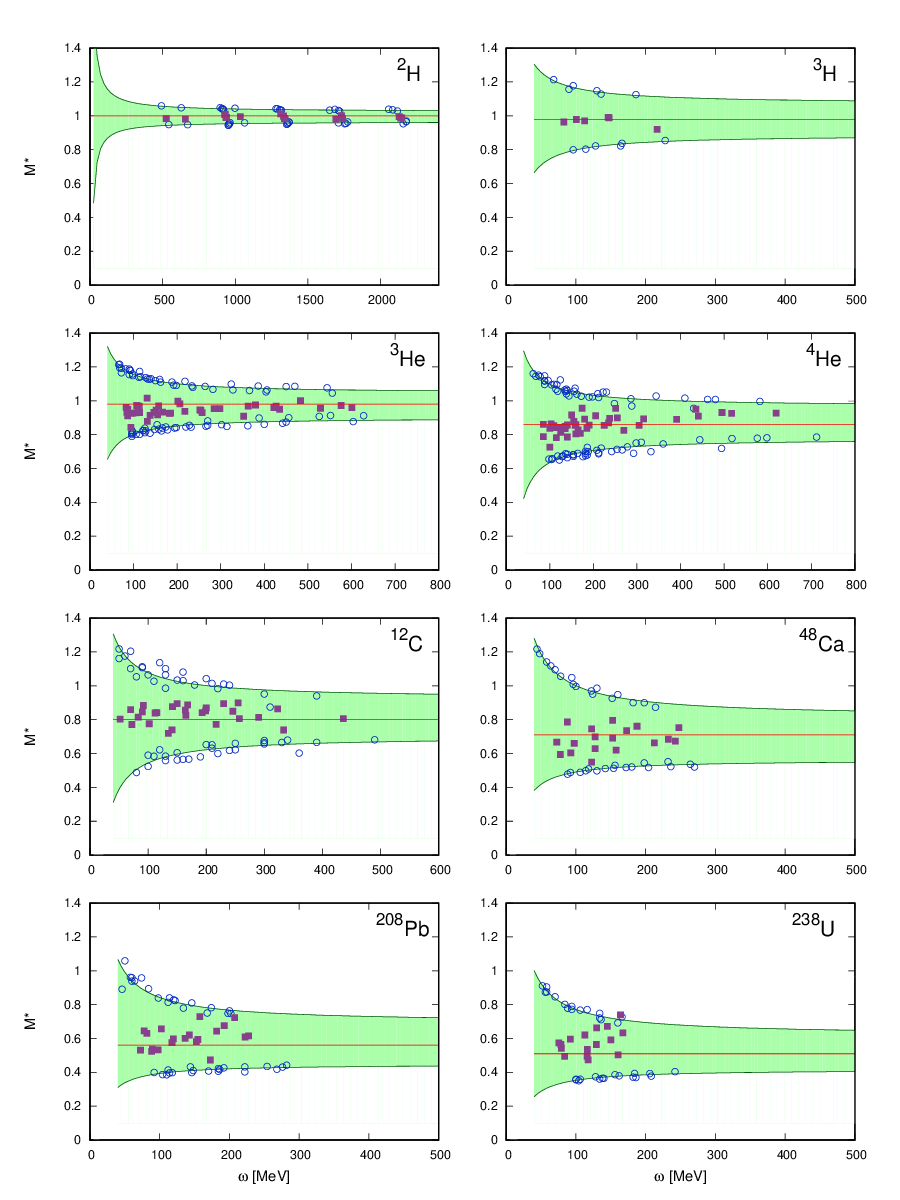}
\caption{Relativistic effective mass $M^*$ computed from the QE peak
  position of the selected experimental data sets (violet squares) as a
  function of the energy transfer $\omega$. The experimental bands
for $M^*$   have been obtained from the SuSAM* bands for the same experimental
  kinematics (circles) by computing the  lower and upper limits for the 
effective mass allowed by the band. 
}
\label{mass}
\end{figure*}

\begin{figure}
\includegraphics[width=9cm]{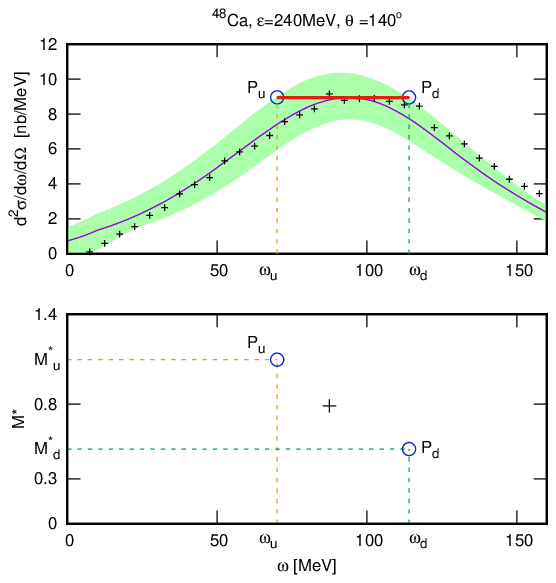}
\caption{Relativistic effective mass $M^*$ computed from the QE peak
  position of the selected experimental data sets (squares) as a
  function of the energy transfer $\omega$. The experimental bands
for $M^*$   have been obtained from the SuSAM* bands for the same experimental
  kinematics (blue circles) by computing the  lower and upper limits for the 
effective mass allowed by the band. 
}
\label{mass2}
\end{figure}

In this section we use the phenomenological scaling function of Eq.
(\ref{scaling-function}) and the parameters corresponding to band B of
table \ref{bandas}, to compute the $(e,e')$ cross section, using the Fermi
momenta and effective masses of columns 6 and 7 of table \ref{tabla}.
 The SuSAM*
model predicts a central cross section inside a theoretical
uncertainty band. The width of the cross section band is related to the width
of our parameterized scaling function for band B. Note that in the
cross section the absolute value of the band width depends on the
kinematics and on the nuclear species.  This is because the cross
section is obtained from the scaling function by multiplication by a
kinematic-dependent function.  Our cross section 
results are here compared to
experimental data for the 21 nuclei included in our fit. This
comparison with the SuSAM* model 
has only been done before for the case of $^{12}$C
\cite{Ama17} and $^{16}$O \cite{Rui18}. We also make comparisons with
the new data for $^{48}$Ti and $^{12}$C performed in a recent
experiment at JLab \cite{Dai18} and provide predictions for the $^{40}$Ar
nucleus corresponding to the kinematics of interest for the JLab
experiment, that plans to extract the Argon spectral function.

The present results have also been studied by using bands A and C and
parameters $k_F,M^*$ from the different fits described in the last
section. The global behavior of the result and the conclusions of
this work are roughly preserved by using any of the three parameterized
scaling functions and bands.

In figs. \ref{H2}--\ref{U238} we show the predictions of our model for
the $(e,e')$ cross section compared to selected experimental data for
each nucleus in the database, from $^{2}$H to $^{238}$U.  The global
description is quite acceptable given the simplicity of the SuSAM*
model.  A large subset of data fall inside the uncertainty band. In
fact, most of the data used to perform the fit are inside our
prediction bands by construction.  The data that lie outside our
prediction bands are typically those in the inelastic or deep region
and those corresponding to low momentum transfer, and therefore
breaking $\psi^*$-scaling because they fall outside the quasielastic
region defined in fig. 1. Alternatively, intermediate energy QE data
falling outside our bands may indicate the existence of nuclear
effects beyond the impulse approximation such as meson-exchange currents
(MEC) or breaking the factorization approximation, like strong final
state interactions.

In what follows we discuss in some detail the cross section description 
 for every single nucleus considered in our study.

\subsection{The nucleus $^2$H}

The lighter nucleus considered in the fit corresponds to $^2$H, shown
in Fig.  \ref{H2}. 
We use the $\chi^2$ fit values from Tab. \ref{tabla} for the 
Fermi momentum, $k_F= 82$ MeV/c, and effective mass $M^*=1$.
The fact that the SuSAM* ---based on the Fermi gas
equations--- reproduces a large fraction of deuterium data could seem
shocking. But what our results reflect is that the deuterium
quasielastic cross section is compatible with a momentum distribution
of moderate extension, $ k_F \sim 82$ MeV/c. This behavior dominates
around the quasielastic region $-1 < \psi^* <1$, while we in general
sub-estimate the left tail of the cross section, corresponding to
$\psi^* < -1$, and is related to PN short range correlations.

\subsection{The nuclei $^3$H, and $^3$He}

The $A=3$ light nuclei analyzed, $^3$H and $^3$He, are shown in
Figs. \ref{H3} and \ref{He3}, respectively. They are even better described than
$^2$H, with slightly different Fermi momenta, $k_F = 136$ and 130
MeV/c for H and He, respectively. The effective mass takes the
same value $M^*=0.98$ for both nuclei.  The value of the Fermi
momentum differ from the other fits, see tab. \ref{tabla}, ranging between
120--140 MeV/c.  All these values of $k_F$ give qualitatively similar
predictions for the cross section, because the cross section
dependence on $k_F$ is mild in a small momentum interval. The
dependence on $M^*$ is found to be stronger. However note that the
differences between the adjusted parameters for these nuclei can also
be related to the different number of experimental data, which is much
larger for the case of $^3$He than for $^3$H.

\subsection{The nucleus $^4$He}

The $^4$He nucleus is the nucleus with more quasielastic data in the
database, and is one with the better scaling properties.  Selected
cross section predictions are shown in Fig.  \ref{He4}, computed with
$k_F=180$ MeV/c and $M^*=0.86$, although equally good results can be
obtained with $k_F=160$ MeV/c and $M^*=0.9$. Comparing with the cases
$A=2, 3$ we clearly see that the Fermi momentum increases with $A$,
and the effective mass decreases with $A$. The QE cross section
description is quite good for many kinematics. It is remarkable that,
given its simplicity, for intermediate energy the SuSAM* model seems
to be as good as the more recent ab initio calculations \cite{Roc18}
with relativistic corrections.  In our view, this is so because the
main dynamical ingredients of the QE processes are embedded into our
model via the connection between relativity and effective mass.

\subsection{The nucleus $^{ 12 }$C }

Results for $^{12}$C are shown in Fig. \ref{C12}. We use $k_F= 217$
MeV/c and $M^*=0.8$. Note that the SuSAM* model was first introduced
in Ref. \cite{Ama17}, where a comparison with all $^{12}$C data was
provided using band A of Tab. \ref{bandas} 
and $k_F=225$ MeV/c. This value of
the Fermi momentum was obtained with a visual fit. In this work we
have upgraded this description using band B, which incorporates a
better description for low energy. But in general both descriptions
are globally of the same quality around the QE peak.  This nucleus is,
together with $^4$He, one of the better described nuclei taking into
account the high number of data existing in the data base, $\sim 2800$
each of them.

\subsection{The nuclei 
$^{ 6 }$Li  ,   
$^{9  }$Be  ,   $^{ 24 }$Mg  ,   $^{59  }$Ni  , 
  $^{89  }$Y  ,   $^{ 119 }$Sn  ,   
$^{181  }$Ta  ,   $^{186  }$W,  and   $^{197  }$Au }

For these nuclei there are only a few set of kinematics available in
the database, usually $\sim 20$--40 QE data points only.  Therefore it
is not possible to determine the parameters with high precision by
maximizing the number of points inside the band, because there are
many arrangements of the data points compatible with the theoretical
band.  For this same reason the $\chi^2$ fit to each nuclei usually
provides a good description of the experimental data for intermediate
energy kinematics, where the scaling approach works better.  This is
shown in Fig. \ref{nuclei} with the parameters $k_F$ and $M^*$ taken
from columns 6 and 7 of table II. The QE data from Li to Au all fall
with our band and are in general well described by the central value,
with the exception of Mg and Au, which are slightly underestimated and
overestimated at the peak position, respectively.

\subsection{The nucleus $^{16}$O}

There are only a limited number of kinematics available for
$^{16}$O. However they are enough for performing all the fits, with a
good global description of data.  We obtain $k_F= 250$ MeV/c in the
$\chi^2$ fit and $M^*=0.79$. In the global and visual fit the Fermi
momentum is more similar to that of $^{12}$C. The comparison of band B
with the experimental $(e,e')$ cross section was shown in Fig. 1 of
ref. \cite{Rui18}, where the SuSAM* parameters were first extracted
for this nucleus, in order to apply the SuSAM* model to neutrino
scattering on water, of interest for the recent T2K experiment
\cite{Abe18}. More experimental kinematics for the $(e,e')$ reaction
on $^{16}$O would be needed to reduce the uncertainty of the SuSAM*
parameters.

\subsection{The nucleus $^{27}$Al}

A number of $\sim 100$ QE-like data are available for $^{27}$Al in the
data base. We show  three kinematic sets in Fig. \ref{Al27}.  This
allows to extract $k_F=249$ MeV/c and $M^*=0.8$ in the $\chi^2$ fit.
Even with so limited number of data, it is also possible to extract
these values by maximizing the number of points inside the band,
obtaining similar values $k_F=258$ MeV/c, and $M^*=0.78$. In the global fit
the Fermi momentum is slightly reduced to 233 MeV/c, similar to the
visual fit value.

\subsection{The nuclei $^{40}$Ca and $^{48}$Ca}

More abundant sets of data are available for the calcium isotopes
$^{40}$Ca and $^{48}$Ca, with very similar values of Fermi momenta
$k_F \sim 236$ MeV/c in the $\chi^2$ fit and $M^*=0.8$. In general these
nuclei are well described by the SuSAM* band for intermediate energies
as shown in the selected kinematics of figs. \ref{Ca40} and \ref{Ca48}.
The number of points inside the cross section bands amount to 616 and
728, for slightly larger values of $k_F= 250$ and 242 MeV/c,
respectively, and $M^*=0.73$ and 0.75.

\subsection{The nucleus $^{56}$Fe}

Iron is an important case of study for some of the neutrino
oscillation ongoing or planned experiments. In addition,
 this is another nice example
with abundant electron scattering experimental data for intermediate
energy and this allows a precise determination of $k_F \sim 240$
MeV/c and $M^* \sim 0.7$. Cross section results for this nucleus are shown in
Fig. \ref{Fe56}.  A fair description of the data is obtained
for incident energies $\epsilon <  2$ GeV.

\subsection{The nucleus $^{208}$Pb}

For heavy nuclei the SuSAM* description is still possible even if the
model is based on the factorization of a single nucleon cross section,
which is expected to be violated by a strong final-state interaction
(FSI) on nuclei such as $^{208}$Pb, shown in Fig. \ref{Pb208}.  An
indication is that the value of $\chi^2/N_{QE}$, given in the last
column of Table \ref{tabla}, is bigger than one for all the heavy nuclei with
abundant number of data. In particular, for lead,
$\chi^2/N_{QE}=1.223$. This fit provides a Fermi momentum $k_F= 233$
MeV/c, similar to the values obtained for other medium/light nuclei,
and effective mass $M^*=0.56$. The other fits performed give rather
similar numbers for these parameters. We have tried to include Coulomb
corrections for this nucleus in terms of an effective momentum, but
the results actually worsen. Additional effects beyond the impulse
approximation mainly meson exchange currents, are expected to be
specially important for this nucleus, and should be investigated in
more depth in the future.

\subsection{The nucleus $^{238}$U}

Results for the heaviest nucleus analyzed, $^{238}$U, are shown in
Fig. \ref{U238}. Even if the number of existent QE data is not so big
as for lead, the description of this nucleus is the worst of all the
nuclei analyzed. The $\chi^2/N_{QE}=1.74$ is the largest appearing in
the last column of Table \ref{tabla}, and the Fermi momentum obtained in the
fit is surprisingly low $k_F=219$ MeV/c, compared to lighter
nuclei. Moreover in the global fit a very different value is obtained
$k_F=255$ MeV/c, closer to the expected value of nuclear matter. This
again indicates that for heavy nuclei strong effects breaking the
impulse approximation and the superscaling hypothesis should play an
important role in the QE regime.

\begin{table}
\begin{center}
\begin{tabular}{|c|c|c|c|c|c|} \hline
Nucleus & $A_u$ (MeV) & $B_u$  & $A_d$ (MeV) & $B_d$ & $\Delta M^*$ \\ \hline
$^2$H & 15.984 & 1.024  & -11.655 & 0.965 & 0.03\\ \hline
$^3$H & 9.377 & 1.07 & -8.94 & 0.888 & 0.09 \\ \hline
$^3$He & 11.06 & 1.045 & -9.85 & 0.9 & 0.07 \\ \hline
$^4$He & 13.133 & 0.967  & -14.202 & 0.777 & 0.1\\ \hline
$^{12}$C & 15.236 & 0.925  & -15.562 & 0.700 & 0.11 \\ \hline
$^{27}$Al & 23.636 & 0.989  & -15.044 & 0.735 & 0.13 \\ \hline
$^{40}$Ca & 18.241 & 0.816  & -6.374 & 0.556 & 0.13 \\ \hline
$^{48}$Ca & 18.618 & 0.815  & -7.247 & 0.563 & 0.13 \\ \hline
$^{56}$Fe & 19.57 & 0.828  &  -12.014 & 0.598 & 0.12\\ \hline
$^{208}$Pb & 14.952 & 0.693 & -5.495  & 0.448 & 0.12  \\ \hline
$^{238}$U & 15.316 & 0.618  & -6.479 & 0.418 & 0.1 \\ \hline
\end{tabular}
\end{center}
\caption{Parameters of the effective mass bands for different
  nuclei. The up and down coefficients are shown in columns 2 to
  5. The theoretical error $\Delta M^*$ of column 6 correspond to the
  $\omega \rightarrow \infty$ limit.  }
\label{parameters}
\end{table}

\subsection{$M^*$ uncertainty}

In the previous results we have assumed that the effective mass is a
constant parameter, which is determined by the position of the QE
peak.  While our parameterization of the scaling function in general
describes well the position of the QE peak, we observe deviations for
some kinematics.  In fact it is observed that for large $Q^2$ the QE
peaks shows a shift to high energy.  This is observed in particular,
for $^4$He when $\omega>300$ MeV.  One could try to improve the
description by using a different 'optimal' effective mass for each  kinematics.
In figure \ref{mass} we show this 'optimal' effective mass
computed for eight nuclei for each experimental kinematic set.
Each set is defined by fixed
incident electron energy and scattering angle. This effective mass is 
plotted as a function of
$\omega$ at the QE peak. The optimal effective mass has been
 computed from the maximum
of the experimental cross section by imposing the quasielastic
condition
\begin{equation}
\omega = \frac{|Q^2|}{2 m_N^*}
\end{equation}
from where
\begin{equation}
M^* = \frac{1}{m_N} \left(\frac{|Q^2|}{2 \omega}\right)_{\rm max}
\label{effmass}
\end{equation}

This allows us to estimate a theoretical uncertainty in the effective
mass obtained from the SuSAM* bands by this method, shown as the blue
circles defining the borders of the green uncertainty band in
Fig. \ref{mass}. The circles in Fig. \ref{mass} have been obtained for
each experimental kinematics from our cross section prediction by the
method displayed for example in fig. \ref{mass2} for a kinematics in
$^{48}$Ca. The used method is as follows. We first draw the horizontal
segment crossing the maximum of the central cross section. Then, we
compute the two points $P_u$ and $P_d$ at the upper border of the
band. These points are an estimation of the minimum and maximum
$\omega$ position of the theoretical QE peak allowed by our band.
From these two $\omega_u$ and $\omega_d$ we compute the up, $M_u^*$,
and down, $M^*_d$, values of the effective mass by
Eq. (\ref{effmass}).  This gives us the two values of the 'optimal'
effective mass shown in the bottom panel of Fig. \ref{mass2}, each one at
a different $\omega$ position, $\omega_u$ and $\omega_d$,
respectively. Repeating this procedure for each experimental
kinematics, we have obtained the blue circles in Fig. \ref{mass}.
Finally, to obtain the green bands defining our estimation of the
effective mass uncertainty by this procedure, we fit the borders of
the band to the resulting points using the parameterization
\begin{equation}
M^*_{u,d} = \frac{A_{u,d}}{\omega} + B_{u,d}
\end{equation}
An estimation of the uncertainty in $M^*$ can be obtained for $\omega\rightarrow \infty$ as
\begin{equation}
\Delta M^* = \frac{B_u-B_d}{2}
\end{equation}
The values of the parameters $A_{u,d}$, $B_{u,d}$ and $\Delta M^*$ are
given in table \ref{parameters}.  Their values amount roughly $\sim
0.1 $ except for deuterium. The uncertainties in the effective mass 
are larger for heavier
nuclei.

\subsection{Predictions for $^{48}$Ti and $^{40}$Ar}

\begin{figure*}
\includegraphics[width=\textwidth]{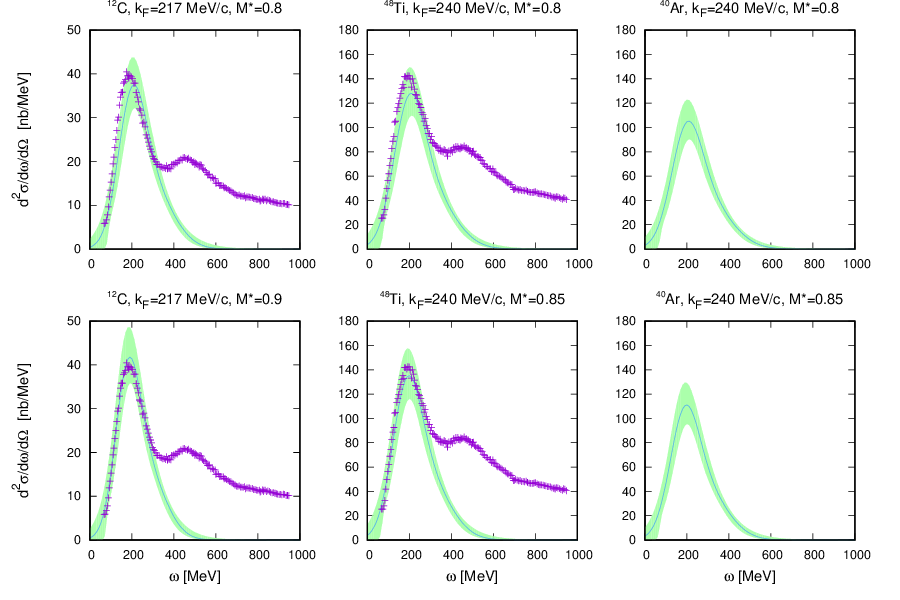}
\caption{Inclusive $(e,e')$ cross sections of $^{12}$C, $^{48}$Ti
and $^{40}$Ar  
for the kinematics $\epsilon=2222$ MeV, $\theta=15.541^{\rm o}$,
of the recent JLab experiment \cite{Dai18},
 compared to the SuSAM* model as a function of 
energy transfer.
}
\label{Ti48}
\end{figure*}

In ref. \cite{Dai18} the first measurement of the $^{48}$Ti$(e,e')$
cross section at Jefferson Lab was reported. The beam energy is
$E=2.222$ GeV and electron scattering angle $\theta=15.541$ deg.  over
a broad range of energy transfer. The purpose of this experiment was
to obtain accurate quasielastic cross section data for the nuclei Ti
($Z=22$) and Ar ($N=22$) to extract information needed for the
neutrino experiments. In the electron
scattering  experiment the cross section of $^{12}$C
for the same kinematics was also measured for calibration. These data
are compared with the SuSAM* model in Fig. \ref{Ti48}. In the upper
panels we show our predictions using the parameter values from the
previous fit, i.e., $k_F \sim 217$ MeV/c for $A=12$ and 
$k_F \sim 240$ MeV/c for
$A=48$, and $M^* = 0.8$ for both nuclei. We observe that our central
prediction is slightly shifted towards high energy transfer with
respect to the data for both nuclei. This means that for this
particular kinematics a higher value for the effective mass is
favored. In fact the $^{12}$C data are better described using
$M^*=0.9$, and $^{48}$Ti needs $M^*=0.85$ (see lower panels of
Fig. \ref{Ti48}). This is related again to the fact that the RMF model
with a constant effective mass starts to fail for high momentum
transfer, favoring an energy dependence of the effective mass, as
mentioned in the previous subsection. The values of Fermi momenta are
similar to those used in the analysis performed in \cite{Dai18} for
this kinematics.

To finish we show in the last column of Fig. \ref{Ti48} our
predictions for the $^{40}$Ar$(e,e')$ quasielastic cross section for
the same kinematics, of interest for the new experiments being
performed at JLab \cite{Dai18,Pan17,Ben14}. For argon we show our band
prediction for $k_F=240$ MeV/c and for two possible values of the
effective mass for this kinematics, $M^*=0.8$ and 0.85. The knowledge
of the experimental cross section for this kinematics would be useful in
our formalism to extract the precise value of the effective mass,
needed to describe the neutrino cross section for the same kinematics.
Note that, although a previous experiment on Ar \cite{Ang95,Ank08} there exist,
the momentum transfer is too low for a reasonable simultaneous extraction
of the effective mass and the Fermi momentum with our formalism.

\section{Conclusions}

In this paper we have analyzed the world data of inclusive quasielastic
electron scattering within the SuSAM* model.  This is a novel scaling
approach based on the relativistic mean field model of nuclear matter
instead of the more usual non-interacting relativistic Fermi gas.  The
new scaling variable $\psi^*$ thus incorporates dynamical ingredients
through the relativistic effective mass $M^*$ which emerges from the
scalar and vector potentials in the Walecka model.  We have applied
several methods to obtain a phenomenological scaling function
$f^*(\psi^*)$ from the inclusive $(e,e')$ reaction data. Our
procedure is to start from a scaling function extracted from  $^{12}$C
data as initial guess, 
and use it to extract the effective mass and Fermi momentum of the
remaining nuclei. In this work we have checked that this method is
consistent with performing a global fit of the scaling function and
all the parameters $k_F,M^*$ over the full database for 21 nuclei.

Thus, superscaling has been shown to be valid for a large body of the
$(e,e')$ data, because the experimental scaling function $f^*(\psi^*)$
collapses into a  thick band that here has been parameterized with
combinations of simple Gaussian functions.  This unique function
allows to describe the intermediate energy QE cross section for light
to heavy nuclei, from $^2$H to $^{238}$U, with tabulated values of
$k_F$ and $M^*$.  Our fit has also allowed to estimate the error in
the extracted parameters $\Delta k_F\sim 10$ MeV/c and $\Delta M^*\sim
0.1$.

Three similar parameterizations, A, B, C, of the phenomenological
scaling function $f^*(\psi^*)$ and uncertainty band have been
tabulated. With these we have presented a systematic analysis of the
predicted QE cross sections and uncertainties  compared to the
data. We observe that the uncertainty band thickness depends on the
kinematics. More than 9000 data of the total 20000 data are found to
be 'quasielastic' as they fall inside the uncertainty band.  The
present results have been shown using band B, but they have also 
been studied by using bands A and C and
parameters $k_F,M^*$ from the different fits. The global results and
the conclusions of this work are preserved against these alternative
parameterizations.

Our model provides one of the best global descriptions of QE data with
a single nuclear model. The success is due not only because its
parameters have been obtained by fitting the inclusive cross section
directly. One crucial reason for the good results is because our model
contains by construction the enhancement of the transverse
components of the electromagnetic current. This is due to the
dynamical enhancement of the lower components of the Dirac spinors by
effect of the relativistic effective mass in nuclei, which is lighter
than in free space.

Note that this enhancement of the transverse response is not observed
in the nuclear isoscalar magnetic moments because there is a
cancellation with the RPA corrections in the $\sigma-\omega$
model for low energy and momentum transfers 
\cite{Ich87}. These RPA-type vertex corrections were found to reduce 
the form factors for $q < 1 $ fm$^{-1} \simeq 200$ MeV/c, but gave
rise to an enhancement for larger $q$ \cite{Ich88}. This is consistent
with the assumed fact that the RPA corrections are small for high
$q$. Since our description of the $(e,e')$ reaction is focused for
large $q \geq 2 k_F$ at the QE peak, defined by $\omega=
|Q^2|/2m_N^*$, the quasifree processes are dominant and
collective excitations are expected to play a minor role.

Our model only requires the assumptions of gauge invariance,
relativity and scaling, which determines the values of the
relativistic effective mass and the Fermi momentum.  The model is
blind to the sorts of nuclear effects involved in the quasielastic
interaction, which are encoded into the scaling function $f^*(\psi^*)$
and its uncertainty band. Whatever nuclear effect which breaks the
impulse approximation or the factorization of the cross section on
which scaling is based, is included only on the average. These may
include MEC and FSI as more direct candidates, but also short range
correlations, which should be more important for negative values of
the scaling variable $\psi^* < -1$, out of the range of the fit made
here.  These very same effects probably give rise to many of the
experimental data falling outside our bands. In future work we expect
to reduce the band thickness adding to the scaling model a
contribution from MEC in the 2p-2h channel, which explicitly accounts
for specific scaling violations.

Our scaling function parameterization also provides a simple test for
theoretical studies, which should fall inside the SuSAM* uncertainty
band.  The universality of the scaling function 
allows this model to be extended to provide tight constraints in
quasielastic neutrino scattering for a wide variety of targets. 
In particular we have provided predictions for the $(e,e')$
cross section of Argon, of interest to current and upcoming 
neutrino experiments.

\section{Acknowledgments}

This work has been partially supported by the Spanish Ministerio de
Economia y Competitividad (grants Nos. FIS2014-59386-P and
FIS2017-85053-C2-1-P) and by the Junta de Andalucia (grant
No. FQM-225). V.L.M.C.  acknowledges a contract with Universidad de
Granada funded by Junta de Andalucia and Fondo Social Europeo.


\end{document}